\documentclass[a4paper,twocolumn,11pt,accepted=2019-12-01]{quantumarticle}
\pdfoutput=1
\usepackage[utf8]{inputenc}
\usepackage[english]{babel}
\usepackage[T1]{fontenc}
\usepackage{physics,
    booktabs,
    mathtools, %for \begin{rcases} \end{rcases}
    placeins, %To include \FloatBarrier
    hyperref, %To use put in url, enable links in bibiliography
    amsmath
}
\usepackage[numbers,sort&compress]{natbib}
\usepackage[dvipsnames]{xcolor}
\hypersetup{
    colorlinks,
    linkcolor={blue!50!black},
    citecolor={blue!50!black},
    urlcolor={blue!80!black}
}% remove the box around citation link and bibilio link

\usepackage[caption=false]{subfig} %The caption part is not compatible with revtex
\usepackage{tikz}
\usetikzlibrary{quantikz}

\usepackage{textcomp} %\textmu s

\newcommand{\addressOxMat}{\affiliation{Department of Materials, University of Oxford, Oxford, OX1 3PH, United Kingdom}}
\newcommand{\addressLCN}{\affiliation{London Centre for Nanotechnology, UCL, 17-19 Gordon St, London, WC1H 0AH, United Kingdom}}
\newcommand{\addressUCLEEE}{\affiliation{Dept.\ of Electronic and Electrical Engineering, UCL, London, WC1E 7JE, United Kingdom}}
\newcommand{\addressQMT}{\affiliation{Quantum Motion Technologies Ltd, Nexus, Discovery Way, Leeds, West Yorkshire, LS2 3AA, United Kingdom}}

\begin{document}

\title{A Silicon Surface Code Architecture Resilient Against Leakage Errors}% Force line breaks with \\

\author{Zhenyu Cai} \email{zhenyu.cai@materials.ox.ac.uk} \orcid{0000-0001-5659-4301}\addressQMT\addressOxMat 
\author{Michael A. Fogarty} \addressQMT\addressLCN 
\author{Simon Schaal} \orcid{0000-0003-2573-7300}\addressLCN
\author{Sofia Patom{\"a}ki} \addressQMT\addressLCN 
\author{Simon C. Benjamin} \orcid{0000-0002-7766-5348}\addressQMT\addressOxMat
\author{John J. L. Morton} 
\email{jjl.morton@ucl.ac.uk}
\addressQMT\addressLCN \addressUCLEEE

\maketitle

\begin{abstract}
    Spin qubits in silicon quantum dots are one of the most promising building blocks for large scale quantum computers thanks to their high qubit density and compatibility with the existing semiconductor technologies. High  fidelity single-qubit  gates  exceeding  the  threshold  of  error  correction codes like the surface code have been demonstrated, while two-qubit gates have reached 98\% fidelity and are improving rapidly. However, there are other types of error --- such as charge leakage and propagation --- that may occur in quantum dot arrays and which cannot be corrected by quantum error correction codes, making them potentially damaging even when their probability is  small. We propose a surface code architecture for silicon quantum dot spin qubits that is robust against leakage errors by incorporating multi-electron mediator dots. Charge leakage in the qubit dots is transferred to the mediator dots via charge relaxation processes and then removed using charge reservoirs attached to the mediators. A stabiliser-check cycle, optimised for our hardware,  then removes the correlations between the residual physical errors. Through simulations we obtain the surface code threshold for the charge leakage errors and show that in our architecture the damage due to charge leakage errors is reduced to a similar level to that of the usual depolarising gate noise. Spin leakage errors in our architecture are constrained to only ancilla qubits and can be removed during quantum error correction via reinitialisations of ancillae, which ensure the robustness of our architecture against spin leakage as well.  Our use of an elongated mediator dots creates spaces throughout the  quantum dot array for charge reservoirs, measuring devices and control gates, providing the scalability in the design. 
\end{abstract}

\maketitle

\section{Introduction}

Universal quantum computers promise speed-up in crucial areas like simulation of materials and molecules~\cite{feynmanSimulatingPhysicsComputers1982}, search~\cite{groverFastQuantumMechanical1996, groverQuantumMechanicsHelps1997} and sampling~\cite{temmeQuantumMetropolisSampling2011, montanaroashleyQuantumSpeedupMonte2015}, 
yet they all require high-precision control of quantum states. Quantum error correction codes allow us to trade qubit number for precision in controlling quantum states (mitigating both control errors and natural decoherence), with the surface code being particularly attractive due to its 2D structure, local checking operations and high error threshold close to 1\%~\cite{wangSurfaceCodeQuantum2011}. Surface code architectures have been proposed for leading quantum information processing platforms including superconducting qubits~\cite{fowlerSurfaceCodesPractical2012},  trapped ions~\cite{lekitschBlueprintMicrowaveTrapped2017} and semiconductor spin qubits~\cite{hillSurfaceCodeQuantum2015, ogormanSiliconbasedSurfaceCode2016}. However, the qubit overheads can be significant: it is estimated that $>2\times 10^8$ physical qubits with gate error rate $10^{-3}$ might be needed to perform a non-trivial Shor's factoring algorithm using surface codes~\cite{ogormanQuantumComputationRealistic2017}.
These considerations motivate the development of qubit implementations which offer the prospect for high-density 2D arrays. The high-qubit density offered by silicon-based spin (SS) qubits (as high as $10^9$~cm$^{-2}$) combined with the possibility of leveraging the conventional semiconductor integrated circuit industry~\cite{vandersypenInterfacingSpinQubits2017} make this platform attractive for fault-tolerant universal quantum computing.

Like all qubit hardware approaches, scaling up SS qubits brings a number of practical requirements associated with qubit addressing for calibration, tuning, operation and readout. Indeed, the high qubit densities offered by SS qubits leads to challenges in routing classical control lines, while minimising cross-talk and managing heat dissipation~\cite{vandersypenInterfacingSpinQubits2017}.
A number of architectures for scaling up SS qubit arrays have been proposed to address such challenges: for example, Veldhorst~\textit{et al.}~\cite{veldhorstSiliconCMOSArchitecture2017} proposed a compact quantum dot array controlled via a crossbar geometry, enabling $N$ qubits to be controlled with $\sqrt{N}$ classical control lines, albeit using control transistors below the dimensions of current technology~\cite{vandersypenInterfacingSpinQubits2017}. Li~\textit{et al.}~\cite{liCrossbarNetworkSilicon2018} went further with a half-filled crossbar architecture that provides more space for classical control lines, though the use of shared control lines brings tight requirements for qubit homogeneity and limitations on the parallelisability of operations. Buonacorsi~\textit{et al.}~\cite{buonacorsiNetworkArchitectureTopological2019} have suggested connecting many small quantum dot modules using electron shuttling in order to provide the space for individual control lines. Smaller quantum dot modules are also easier to calibrate and the operations within the modules may be expected to have higher fidelities. However, such shuttling architectures require distribution of entanglement between modules and this is likely to impact the fidelity and speed of inter-module operations.

While such influential architectures have been designed to accommodate error correcting codes that compensate for computational errors, they do not address so-called `leakage errors' in which the quantum system escapes out of the computational subspace. For SS qubits, one form of leakage errors arises from the migration of charge: Controlling SS qubits involves tuning tunnelling barriers, changing on-site energies and/or shuttling electrons, and each of these operations may lead to electrons escaping out of the quantum dots. Since leakage errors of this kind cannot be corrected (and may even be exacerbated) by the usual quantum error correction protocols, they will accumulate and eventually corrupt the surface code even if the probability of these leakage errors is very small. Furthermore, unlike most of the other types of leakage errors~\cite{motzoiSimplePulsesElimination2009, ferronIntrinsicLeakageJosephson2010, duanGeometricManipulationTrapped2001, haffnerQuantumComputingTrapped2008, fongUniversalQuantumComputation2011, mehlFaulttolerantQuantumComputation2015} which occur as independent events, a leaked charge from one dot might propagate through the quantum dot surface code array and corrupt other dots. Charge leakage errors thus could be very damaging to the surface code due to the correlations in errors.

In this Article, we introduce a surface code architecture based on SS qubits that is designed to be robust against leakage errors. We first  introduce the components of our hardware in Section~\ref{sect:physical_components}, and then discuss leakage errors in our architecture in Section~\ref{sect:leakage}. Then, in Section~\ref{sect:surface_code}, we describe how surface code stabiliser checks are performed, and obtain a threshold for the gate errors and leakage errors. Finally, we summarise the key features of this approach and discuss possible improvements and extensions.

\section{Physical Implementation}\label{sect:physical_components}
\begin{figure}[htbp]
    \centering
    \subfloat[]{\includegraphics[width = 0.5\textwidth]{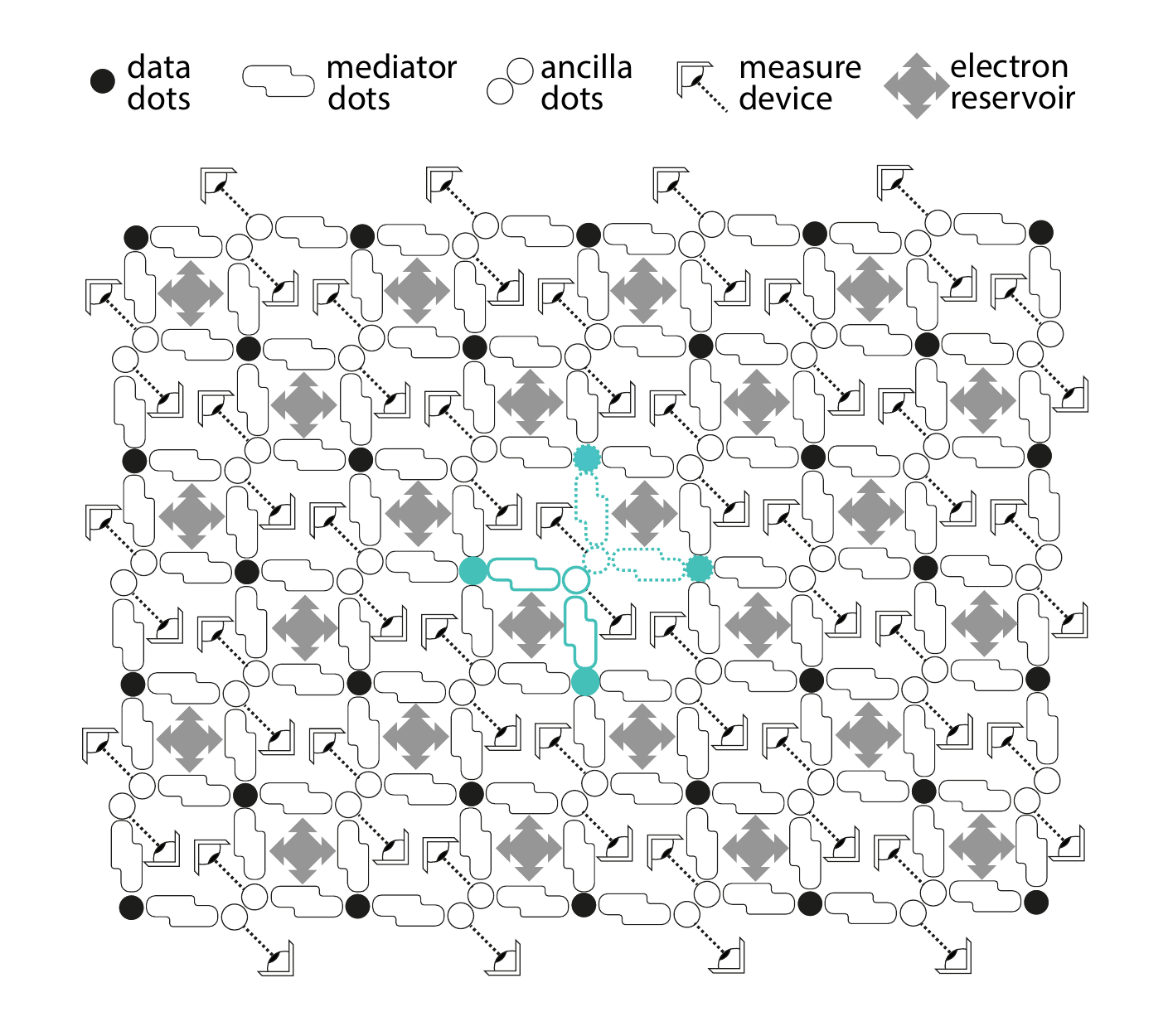}}\\
    \subfloat[]{\includegraphics[width = 0.51\textwidth]{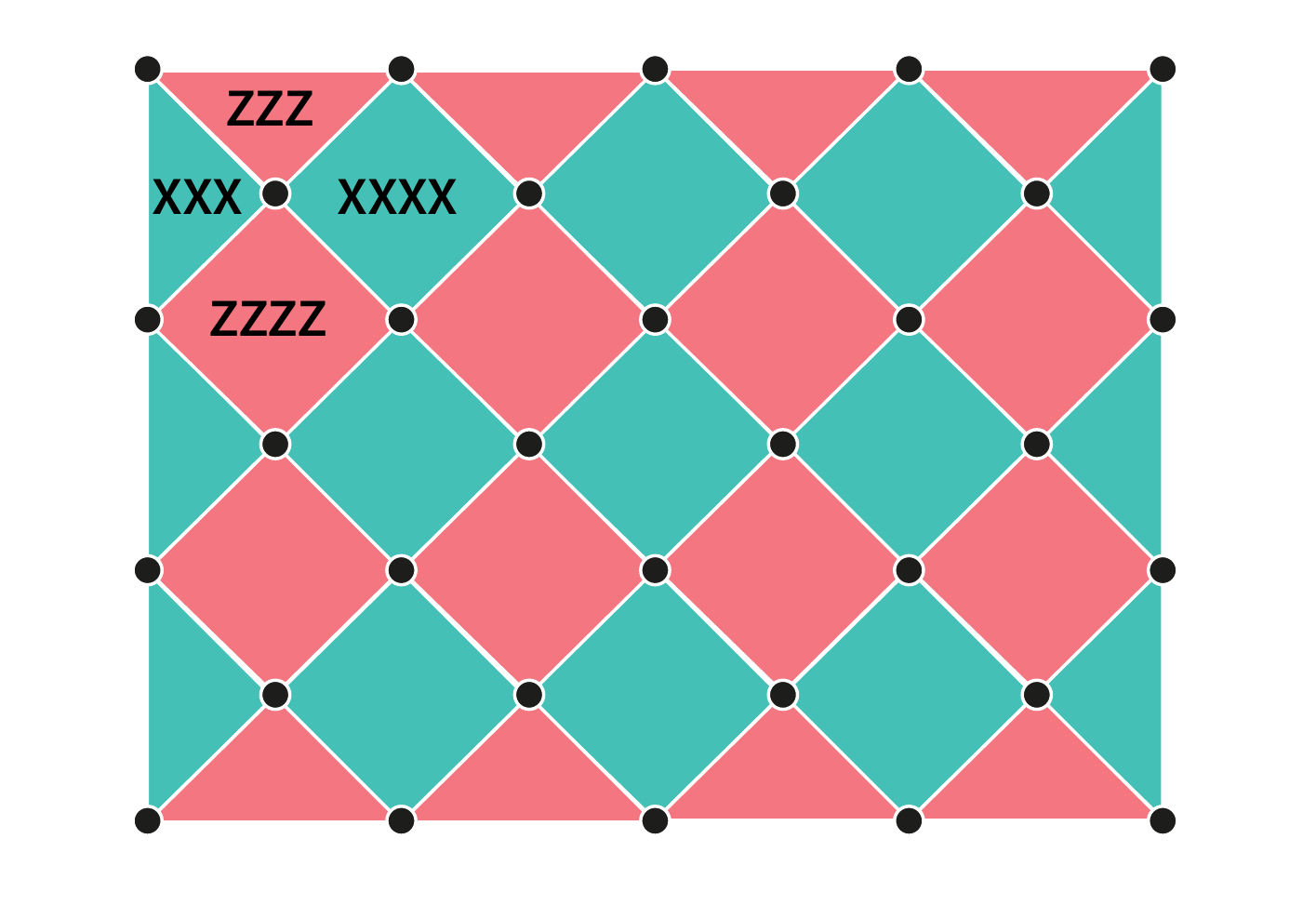}}
    \caption{{\bf Overall architecture layout}, including (a) the arrangement of the key physical components of the system and (b) its correspondence to the components of the surface code. An X-stabiliser plaquette is highlighted in (a), which can be divided into two parts each interacting with one half of the double quantum dot ancilla in the centre.}
    \label{Fig:layout_schematics}
\end{figure}

The physical layout of the silicon quantum dot surface code architecture we consider is shown in Figure~\ref{Fig:layout_schematics}. We have included elongated mediator dots~\cite{malinowskiFastSpinExchange2019} to provide the basic two-qubit gate operation while increasing the fundamental inter-qubit spacing to more readily accommodate measuring devices for ancilla readout, and electron reservoirs for initialisation and reset of quantum dots. Quantum information resides in the \emph{data dots}, whose error information is extracted by the \emph{ancilla double-dots} via  interactions through the \emph{mediator dots}.

\subsection{Data Qubits and Single-qubit Gates}\label{sect:one_qubit_gate}
Each data qubit is represented by the spin state of an electron within an  electrostatically-defined quantum dot~\cite{angusGateDefinedQuantumDots2007}. The lifting of the spin degeneracy via an applied magnetic field gives access to electron-spin resonance (ESR)~\cite{veldhorstAddressableQuantumDot2014,chanAssessmentSiliconQuantum2018, yangSiliconQubitFidelities2019} or electrically-driven spin resonance (EDSR)~\cite{kawakamiGateFidelityCoherence2016,yonedaQuantumdotSpinQubit2018} techniques, which have been used to produce control fidelities of electron spin qubits in silicon of up to 99.6--99.9\%~\cite{veldhorstAddressableQuantumDot2014,yonedaQuantumdotSpinQubit2018}. As has been considered in several proposals~\cite{picaSurfaceCodeArchitecture2016,jonesLogicalQubitLinear2018,veldhorstSiliconCMOSArchitecture2017,liCrossbarNetworkSilicon2018}, driving fields can be applied globally to all, or many, qubits~\cite{tyryshkinElectronSpinCoherence2012,lauchtElectricallyControllingSinglespin2015} in order to avoid the problem of `frequency crowding'~\cite{risteDetectingBitflipErrors2015,schutjensSinglequbitGatesFrequencycrowded2013} when attempting to select many individual resonances within a finite bandwidth. Qubit relaxation times ($T_1$) can reach $>1\, $s~\cite{yangSpinvalleyLifetimesSilicon2013,kawakamiElectricalControlLonglived2014, leonCoherentSpinControl2019} and using isotopically enriched $^{28}$Si substrates~\cite{itohIsotopeEngineeringSilicon2014}, qubit coherence times can be extended up to the limits of the fluctuation timescales within the magnetic environment (e.g. $T_2^* \sim $120\, \textmu s~\cite{veldhorstAddressableQuantumDot2014}). Decoupling schemes can then be used to yield longer qubit operation times ($T_2 \sim$ 28~ms~\cite{veldhorstTwoqubitLogicGate2015}), and these can be integrated into algorithms~\cite{jonesLogicalQubitLinear2018,witzelMultiqubitGatesProtected2015} or single qubit gates designed via gradient ascent pulsed engineering~\cite{khanejaOptimalControlCoupled2005} to be inherently robust against environmental noise~\cite{yangSiliconQubitFidelities2019}. 

Qubits formed electrostatically in highly strained silicon also have the advantages of splitting off the excited states when quantum dots are strongly confined. Such systems often show valley excited states of $\sim0.1\, $THz~\cite{yangOrbitalValleyState2012,yangSpinvalleyLifetimesSilicon2013,veldhorstAddressableQuantumDot2014}, and orbital energies of $\sim 1\, $THz~\cite{yangOrbitalValleyState2012}, and such excited state energies can be electrostatically tuned via the Stark shift~\cite{yangSpinvalleyLifetimesSilicon2013,veldhorstTwoqubitLogicGate2015}. In a confined quantum dot with diameter, say $30\, \text{nm}$, a large Coulomb repulsion $U\sim 2\, $THz~\cite{yangOrbitalValleyState2012} is produced --- this effectively prevents additional charges entering such dots during the execution of the code, leaving us to address the possibility of charge leakage out of the dot.

\subsection{Ancilla Qubits and Read-out}\label{sect:anc}
Our proposed ancilla qubit is represented by the spin state of a \emph{pair} of electrons distributed across two quantum dots (each similar in size to the data dots). By initialising in a singlet state, a failed stabiliser check of its neighbouring data qubits transforms the ancilla spins into a triplet state~\cite{jonesLogicalQubitLinear2018}, such that we can use Pauli spin blockade (PSB) and its effect on interdot tunnelling~\cite{onoCurrentRectificationPauli2002,pettaCoherentManipulationCoupled2005}, to determine the outcome of the stabiliser cycle. PSB can be detected in single-shot through charge sensing~\cite{johnsonSinglettripletSpinBlockade2005}, or via gate-based dispersive readout~\cite{betzDispersivelyDetectedPauli2015, westGatebasedSingleshotReadout2019, pakkiamSingleShotSingleGateRf2018} as suggested by the measurement devices in Figure~\ref{Fig:layout_schematics}~\cite{johnsonSinglettripletSpinBlockade2005, yangDynamicallyControlledCharge2011}. The ancilla qubits are initialised via the (0,2) electron occupation state of the double quantum dot (or an equivalent ($N$, $N+2$) state), where the ground state is a singlet and can be rapidly prepared through `hot-spot' relaxation near the (1,1):(0,2) charge transition~\cite{johnsonTripletSingletSpin2005}.

Previous schemes~\cite{veldhorstSiliconCMOSArchitecture2017, jonesLogicalQubitLinear2018} have employed a second quantum dot as part of the ancilla structure as a reference state which does not participate in the stabiliser check --- the primary function being to enable measurement by PSB. In contrast, in our proposal we treat both ancilla dots on an equal footing, allowing both dots in the ancilla pair to interact with data qubits. In addition to reducing complexity in connectivity, this approach enables interactions between the data qubits and ancillae to be performed in parallel using both of the ancilla dots, halving the time needed to perform a stabiliser cycle. 

Operations that are symmetric under the exchange of the two spins cannot bring the quantum state out of the singlet (exchange-antisymmetric) or the triplet (exchange-symmetric) subspace. Hence, global ESR (single qubit gates) can be applied to all the data qubits without affecting the double-dot ancilla, which is useful when switching between $X$ and $Z$ stabiliser check cycles of the surface code.

The type of error (e.g.\ $X$ or $Z$) detected by single-dot ancillae depends on the basis in which they are prepared and measured, while two-dot ancillae prepared in the singlet state can be used to detect both $X$ and $Z$ errors~\cite{jonesLogicalQubitLinear2018}. In standard parity check circuits, the $X$ and $Z$ errors in the data qubits will be transformed into $Z$ errors in the ancilla qubits using CZ and CNOT respectively, which can be detected by preparing and measuring the ancilla in the X-basis. In the case of double-dot ancilla, this can be achieved by mapping the singlet state and the zero-spin triplet state of the two-dot ancillae to the X-basis eigenstates:
\begin{align*}
\frac{1}{\sqrt{\sqrt{2}}} \left(\ket{01} \mp \ket{10}\right) \mapsto \ket{\pm}_{anc}.
\end{align*} 
On the L.H.S. we have the state of the two physical spins within the ancilla double-dot and on the R.H.S. we have the corresponding state of the ancilla qubit.

From the mapping we can see that while $Z$ gates on individual physical spin correspond to $Z$ gates on the ancilla qubit, $X$ and $Y$ gates on the individual physical spin will take the spin pairs out of the zero-spin subspace, resulting in leakage errors on the ancilla qubit. The effect of such leakage errors will be detailed in Section~\ref{sect:leakage}.

\subsection{Mediators and Two-qubit Gates}\label{sect:two_qubit_gate}
\begin{figure}[t]
    \centering
    \includegraphics[scale = 0.7]{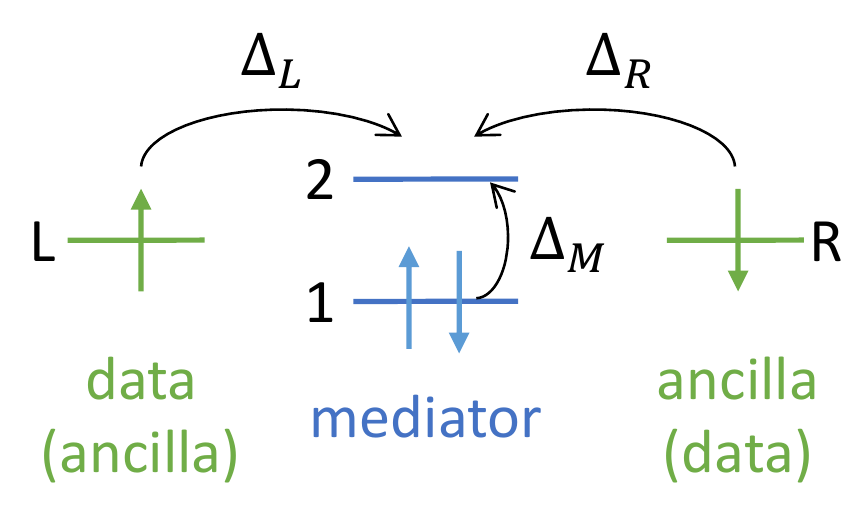}
    \caption{{\bf Core two-qubit gate between data and ancilla qubits, achieved via a mediator}. Three quantum dots with orbital $L/R$ in the left/right dot, each of which is either a data dot or one half of an ancilla structure, and orbitals $1$ and $2$ in the middle mediator dot. We consider a total of four electrons in this three-dot system and assume the charging energy of the side dots is sufficiently large (due to their small size) to forbid further occupancy. Electrons may be excited to the mediator state $2$ from any of $L$,$R$ or $1$ orbitals, with some energy cost indicated.}
    \label{Fig:three_dot}
\end{figure}

When two-qubit gates are performed using direct exchange interactions between nearest-neighbour quantum dots, the resulting qubit pitch is typically on the scale of tens of nanometres. On the other hand,  control and read-out electronics associated with each qubit are more comfortably accommodated with larger spacings at the level of at least several hundreds of nanometres.  To extend the range of the exchange interaction, an elongated quantum dot can be used as a `mediator'~\cite{srinivasaTunableSpinQubitCoupling2015, mehlTwoqubitCouplingsSinglettriplet2014}. Our architecture employs effective two-electron (i.e.\ even-occupation) quantum dots as mediators for the exchange interaction between a data dot and one half of the ancilla double-dot as shown in Figure~\ref{Fig:three_dot}. For simplicity we assume two-electron occupation in the mediator, but in practice four-electron or other values may be preferable, for example to mitigate a small valley-orbit splitting~\cite{harvey-collardCoherentCouplingQuantum2017}.

The mediators do not themselves carry any quantum information and our computational subspace only consists of the spin states of the electrons in the two side dots. Ruderman-Kittel-Kasuya-Yosida (RKKY) exchange interactions communicated by the mediators occur between the spins in the two side dots, with a strength~\cite{srinivasaTunableSpinQubitCoupling2015} given by:

\begin{align}\label{eqn:exchange_strength}
J = -2\left(\frac{t^*_{R2} t_{R1} t^*_{L1} t_{L2} }{\Delta_R\Delta_M\Delta_L} + \text{c.c.}\right) 
\end{align}
where $t_{ab}$ is the tunnelling energy from orbital $a$ to $b$ and $\Delta_{R,M,L}$ are the energies associated with various electron hopping processes starting from the ground state as indicated in Figure~\ref{Fig:three_dot}.

We consider mediator dots of dimensions $30~\text{nm} \times 300~\text{nm}$, which leads to $\Delta_M \sim 10\, $GHz, and assume a tunnelling energy between the mediators and data/ancilla dots of $t \sim 1\, $GHz (corresponding to an interdot spacing of $\sim10\, $nm). By tuning the on-site energy of the mediator dot, we can change the value of $\Delta_{R/L}$ and hence control the strength of the exchange interaction. $\Delta_{R/L}$ is bounded to be at least the tunnelling energy and at most the Coulomb repulsion energy in the data dots. Hence, we use $\Delta_{R/L} = \Delta_\text{\rm on} = 10\, \text{GHz}$ to turn on the exchange interaction, and $\Delta_{R/L} = \Delta_\text{\rm off} = 1\, \text{THz}$ to turn off the exchange interaction. Using (\ref{eqn:exchange_strength}), the strength of the exchange interaction is $J_\text{\rm on} = \frac{t^4}{\Delta_{\rm on}^2\Delta_M} = 1\, \text{MHz}$ when on, and has a residual value $J_{\rm off} = \frac{t^4}{\Delta_{\rm off}^2\Delta_M} = 100\, \text{Hz}$ when nominally off. This level of residual exchange interaction leads to an expected error probability ($\frac{J_{\rm off}}{J_{\rm on}} \approx 10^{-4}$) well below the threshold of the surface codes and hence ignored in our discussion. We assume the mediated exchange is controlled through the detuning of the mediator dot under the fixed tunnel coupling naturally formed between adjacent dots~\cite{veldhorstTwoqubitLogicGate2015} --- it is also possible to use additional electrodes for controlling the tunnel coupling between adjacent dots~\cite{reedReducedSensitivityCharge2016}, albeit at the cost of greater gate complexity. 

A difference in $g$-factors or in $z$-magnetic field in the left and right (L/R) dots produces a difference in the Zeeman splitting between them, which we denote $\Omega$. When the device is tuned to satisfy $\Omega \ll J$, the exchange interaction enables us to implement $\sqrt{\text{SWAP}}$ gates (see Appendix~\ref{app:CZ} for details). Along with single-qubit Z rotations, they can be used to create CZ gates as proposed by Loss and DiVincenzo~\cite{lossQuantumComputationQuantum1998}:
\begin{center}
    \begin{tikzcd}
        &\ctrl{1} &\qw \\
        &&\\
        &\ctrl{-1}&\qw
    \end{tikzcd}
    \begin{tikzcd}
        \\
        \equiv\\
    \end{tikzcd}
    \begin{tikzcd}
        &\gate{Z_{\frac{\pi}{2}}} &\gate[wires=2, style={yshift=-9.5pt, inner ysep=-5pt}]{\rotatebox{270}{$\sqrt{\text{SWAP}}$}}    &\gate{Z_\pi} &\gate[wires=2, style={yshift=-9.5pt, inner ysep=-5pt}]{\rotatebox{270}{$\sqrt{\text{SWAP}}$}}         &\qw\\
        &\gate{Z_{-\frac{\pi}{2}}} &     &\qw&  &\qw\\
    \end{tikzcd}
\end{center}

On the other hand, in the limit where $\Omega \gg J$, we can achieve a dipole-dipole like interaction between the two dots mediated by the exchange interaction, which can be used to implement $S = \frac{1}{\sqrt{2}} \left(I_1I_2 + iZ_1Z_2\right)$ (see Appendix~\ref{app:CZ} for details). Along with single-qubit Z rotations, they can be used to create CZ gates as proposed by Meunier \textit{et al.}~\cite{meunierEfficientControlledphaseGate2011}:
\begin{center}
    \begin{tikzcd}
        &\ctrl{1} &\qw \\
        &&\\
        &\ctrl{-1}&\qw
    \end{tikzcd}
    \begin{tikzcd}
        \\
        \equiv\\
    \end{tikzcd}
    \begin{tikzcd}
        &\gate{Z_{\frac{\pi}{2}}}&\gate[wires=2]{S} &\qw       \\
        &\gate{Z_{\frac{\pi}{2}}}& &\qw 
    \end{tikzcd}
\end{center}

Two-qubit gates in silicon QDs based on direct exchange have been demonstrated~\cite{veldhorstTwoqubitLogicGate2015, watsonProgrammableTwoqubitQuantum2018}, whose fidelity has been improved to 98\%~\cite{huangFidelityBenchmarksTwoqubit2019}, fast approaching the fault-tolerant threshold. Mediated exchange using empty~\cite{baartCoherentSpinexchangeQuantum2017} or multi-electron~\cite{malinowskiFastSpinExchange2019} mediator dots has also been demonstrated in GaAs quantum dots. 
In our architecture, we use an effective two-electron mediator dot to provide exchange interactions that can be more readily to switched on and off than with empty mediator dots due to a lower virtual energy cost, noting also that keeping the occupancy low leads to higher expected fidelity than the multi-electron mediators due to the simpler electron environment in the mediators~\cite{srinivasaTunableSpinQubitCoupling2015}.

\subsection{Realisations of the $\Omega \ll J$ and $\Omega \gg J$ regimes}\label{sect:micromagnet}

As explained above, the RKKY exchange operation produced by the mediator dot can be utilised to construct the CZ operation either directly via the S gate when $\Omega \gg J$, or indirectly via $\sqrt{\text{SWAP}}$ operations when in the $\Omega \ll J$ regime. Embedding our device within a uniformly applied external magnetic field enables single qubit operations via ESR~\cite{veldhorstAddressableQuantumDot2014}, while also accessing the $\Omega \gg J$ regime through the natural variation in electron $g$-factor inherent to the qubit platform~\cite{veldhorstTwoqubitLogicGate2015, ferdousInterfaceinducedSpinorbitInteraction2018}.

However, single-qubit gates achieved by ESR have relatively slow speed ($\sim 1$~MHz), limited by the magnitude of the oscillating magnetic field. An alternative method to implement single-qubit gates is EDSR~\cite{tokuraCoherentSingleElectron2006}, which can be achieved in a uniform magnetic field~\cite{cornaElectricallyDrivenElectron2018} by exploiting the spin-orbit coupling in silicon~\cite{jockSiliconMetaloxidesemiconductorElectron2018,tanttuControllingSpinorbitInteractions2019}.
More commonly, EDSR is achieved using a magnetic field gradient created at the quantum dot, usually by placing a micromagnet in proximity. In this way, when the electron is perturbed via an oscillating electric field, it experiences an effective oscillating magnetic field which drives the spin rotation. EDSR can be more than an order of magnitude faster ($>10$~MHz~\cite{yonedaQuantumdotSpinQubit2018}) than ESR, however, qubits capable of EDSR driving can be more susceptible to decoherence from charge noise of the control gates. A balance can be struck between the speed of the single-qubit gates and qubit decoherence to achieve single-qubit EDSR gates with fidelity of $99.9\%$~\cite{yonedaQuantumdotSpinQubit2018}.

As shown in Section~\ref{sect:anc} and further discussed in Section~\ref{sect:stb_circuits}, we do not need to apply single-qubit gates to our ancillae, and so there is no advantage in furnishing them with micromagnets. This fact, together with the additional spacing between qubits afforded by the mediator dots, facilitates the ability to deposit a micromagnet array such that each data qubit is located in the vicinity of a magnetic field gradient. This local field gradient observed by the data qubits facilitates EDSR, and also produces the offset field between data and ancilla qubits attaining the $\Omega \gg J$ regime.

A second regime in which the qubit array can be operated is the $\Omega \ll J$ regime. In order to achieve this regime, no field gradients due to micromagnets are utilised and the external magnetic field must be low, such that $\Omega$ attributed to the variation in the electron $g$-factor is minimal. Given current disorder levels within Si QDs, the use of an applied field of $\sim$30~mT (enabling ESR at $\sim$1~GHz) would yield in $\Omega \sim J$. To push into the $\Omega \ll J$ regime, the platform could be further engineered for larger $J$ values, or for the reduction in disorder levels giving rise to variation of electron $g$-factors such that they can be mitigated effectively using the Stark shift.

\subsection{Charge Reservoirs and Initialisation}\label{sect:charge_reservoirs}
Charge reservoirs remain an integral component of modern test-bench quantum devices as they are used to supply electrons to quantum dots, facilitate traditional spin-to-charge readout~\cite{elzermanSingleshotReadoutIndividual2004} or more recent improved methods~\cite{harvey-collardHighFidelitySingleShotReadout2018, fogartyIntegratedSiliconQubit2018}, as well as providing a relaxation path for rapid spin initialisation~\cite{johnsonTripletSingletSpin2005}. However, modern concepts of scaled qubit platforms that exploit CMOS technology typically envisage larger devices with densely-packed quantum dots, leading to reservoirs being pushed to the borders of large 1D~\cite{jonesLogicalQubitLinear2018} or 2D~\cite{veldhorstSiliconCMOSArchitecture2017} arrays. Other architectures have the capacity for reservoirs to be located in specialised modules where spins could then be shuttled into arrays through the use of long-distance highways~\cite{liCrossbarNetworkSilicon2018}. With the relative absence of reservoirs in many modern architectures, spin initialisation and readout relies predominantly on Pauli spin blockade methods, with some schemes also utilising thermal relaxation as an initialisation method~\cite{vandersypenInterfacingSpinQubits2017}.

In the architecture presented here, we strive to maintain the advantages of having integrated spin reservoirs, without compromising the advantages of CMOS as a platform capable of realising arrays of densely-packed qubits. This is achieved through the spatial separation afforded by the larger scale mediator dot between each data/ancilla dot as seen in Figure~\ref{Fig:layout_schematics}. With a gate pitch of 30--40~nm~\cite{yangSpinvalleyLifetimesSilicon2013,veldhorstAddressableQuantumDot2014} in recent 2D planar SiMOS QD designs, and with the possibility of reducing this through the use of smaller length scales (e.g.\ more recent CMOS technology nodes), the indicated 300~nm separation due to the mediator generates enough space for the integration of the reservoirs as well as the planar fan-out of metallic gate structures required to define/confine the 2D quantum dot structures. Specifically, this facilitates the ability to maintain gated connections between the reservoir and the mediator dot, meaning the tunnel rate can be tuned or made switchable for either rapid interaction as required during initial population of a qubit array, or appropriately tuned for slow reset of mediator dots during periods of inactivity. 

The smallest energy scale for the mediator system is $\Delta_M \sim 10$~GHz, which remains $\sim 5\times$ larger than conservative electron temperatures of $\sim100$~mK. Couplings required for Elzerman readout~\cite{elzermanSingleshotReadoutIndividual2004} are $\sim100$~\textmu s in typical CMOS systems~\cite{veldhorstAddressableQuantumDot2014}, which is long compared to the CZ execution time, however dispersive sensing has seen device operation with tunnel couplings on the order of tank circuit frequencies of 1--10~MHz, which would place the mediator reset, or initialisation protocols within an appreciable time budget with respect to the error correction scheme. This is made possible because this scheme does not utilise the reservoir for coherent operations such as readout, and hence the tunnel rate can be made larger than timescales required for high fidelity single-shot detection via classical electronics. 

\section{Leakage errors}\label{sect:leakage}
\subsection{Background}
A \textit{leakage error}, in which the state of the quantum system escapes out of the computational subspace, is not corrected by typical quantum error correction protocols. If left uncorrected, even low-probability leakage errors may accumulate and eventually corrupt the logical qubits. Wood and Gambetta have presented an recent overview on leakage error models and how they can be quantified~\cite{woodQuantificationCharacterizationLeakage2018}.
To correct leakage errors, we need to first reduce them to errors that fall within the computational space, which can then be handled by the quantum error correction scheme.
This can be achieved by detecting the leakage errors~\cite{preskillFaulttolerantQuantumComputation1998, gottesmanStabilizerCodesQuantum} and replacing the leaked qubits with fresh qubits, or employing leakage reduction protocols~\cite{aliferisFaulttolerantQuantumComputation2007, fowlerCopingQubitLeakage2013, sucharaLeakageSuppressionToric2015} to all qubits without the need of leakage detection. In practice, the sources of, effects of, and  solutions to leakage errors are strongly hardware-dependent. 

In our architecture, we use the term \emph{qubit dots} to refer both to data dots and ancilla dots in which quantum information resides. Within our computational subspace, all qubit dots will be in the ground charge configuration. The electrons in the data single-dots are allowed to have any spin configurations while the electron pairs in the ancilla double-dots are restricted to the spin-zero subspace. The wrong spin or charge configuration of the system will leads to spin leakage or charge leakage errors respectively.

\subsection{Robustness Against Spin Leakage Errors}\label{sect:robust_against_spin_leakage}
Spin leakage error means the spin configuration of the system go out of the spin subspace that defines the computational subspace, given the right charge configuration. There is no spin leakage for the data qubits in our case since all of their spin configurations are within the computational subspace. Hence, the only spin leakage in our architecture will be the ancilla qubits escaping out of the spin-zero subspace. Ancilla spin leakage cannot spread to the data qubits via interactions (since there is no data spin leakage), which means that spin leakage cannot propagate in our architecture. Furthermore, the spin leakage errors of the ancilla qubits will be removed in every new round of stabiliser checks when we reinitialise the ancilla. These properties ensure spin leakage will not lead to spatially or temporally correlated errors in our architecture, permitting robustness against spin leakage.

Now if we take a look at the stabiliser check circuit in Figure~\ref{Fig:stb_circuit}, we can see that before the readout, the stabiliser check process can be viewed as two non-interacting halves. Within each half, there will be two data qubits interacting with one spin within the ancilla spin pair in the same way as interacting with a single-spin ancilla qubit. Hence, before the readout, we can study all the errors on an ancilla qubit simply by treating each spin within the ancilla spin pair as an individual qubit. The spin leakage errors of the ancilla qubits can be taken into account in this way because they can be represented by unitaries applied on the ancilla spin-pair, e.g. $X$ and $Y$ gates on individual spins are two possible forms of ancilla spin leakage errors as mentioned in Section~\ref{sect:anc}. Hence, we can see that the effect of spin leakage errors on our double-dot ancillae should be similar to the effect of computational errors in some alternative schemes using two single-dot ancillae. 

In the readout stage, we need to consider the errors on both spins together. The double-dot ancilla singlet-triplet readout may fail when there are non-symmetric errors occurring on the two spins, compared to the single-dot ancilla $X$-basis readout which will fail under any non-$X$ errors.

\subsection{Robustness Against Charge Leakage Errors}\label{sect:robust_against_charge_leakage}
Charge leakage error means the charge configuration of the qubit dots moves away from the ground charge configuration that our computational subspace resides in. In our architecture, the electron-electron repulsion energies in the qubit dots are much higher than any other energy in our system, thus we do not consider the charge leakage errors due to extra electrons entering the qubit dots. Instead, we will focus on the charge leakage errors due to electrons escaping out of the qubit dots. Possible sources of such charge leakage errors include decoherence of charge eigenstates during exchange interactions~\cite{barrettDoubleoccupationErrorsInduced2002} (see also Appendix~\ref{app:leakage_mechanism}) or electrons escaping out of the 2D electron gas confinement. 

Charge leakage is much more damaging than spin leakage in two ways. First of all, charge leakage can be transferred from one qubit to another via gate operations, which will lead to propagation of leakage errors in the qubit array. Secondly, it cannot be simply removed by reinitialisation of the spin configuration. The missing charge must be replenished using charge reservoirs, which can be hard to integrate into a densely-packed quantum dot arrays. 

When an electron escapes from a qubit dot in our architecture, it can be restored via relaxation of electrons in the neighbouring mediator dots into the empty qubit dot. The time scale of such relaxation is indicated by the $T_1$ time of charge qubits in semiconductor quantum dots. Wang~\textit{et al.}~\cite{wangChargeRelaxationSingleElectron2013} measured the charge relaxation time in Si/SiGe double quantum dots, showing strong dependence on the tunnelling energy between the orbitals and weak dependence on the detuning between the orbitals. For the tunnelling energy regime that we are interested in ($t \sim$1~GHz), the relaxation time was around $10$~ns, which is much shorter than the other time scales in our systems (all the gates in our system operate at \textmu s time scale). Hence, we can assume that once a charge leakage error occurs, a relaxation process quickly takes place, in which an electron in one of the adjacent mediator dots hops down to fill the empty qubit dot, restoring the charge configuration of the qubit dots. Therefore, even without any active leakage error detection and correction or applications of any leakage reduction protocols, our architecture has a useful inherent behaviour whereby charge deficit transfers from qubit dots to mediator dots.

The relaxation process that restores the charges in the qubit dots can, however, result in missing/extra charges in the mediator dots, which, uncorrected, would produce faulty exchange gates. This can be corrected by connecting all the mediators to the charge reservoirs that are used for the initial population of the quantum dot array. Since the mediators do not carry any quantum information, such connection to reservoirs should not introduce qubit errors. 

Errors due to unwanted coupling between the charge reservoirs and the qubit array are minimised by decreasing the tunnelling energy between the reservoir and the mediators, though this produces a longer reset time for the mediators. As we will see in Section~\ref{sect:surface_code}, our surface code is partitioned into regions which are active/inactive at different times during a full cycle. This provides an opportunity for a given mediator to reset with its nearby reservoir during an idle period, without adding delay to the error correction processes. The tunnel coupling between mediator and reservoir can be minimised to the level required to give a reliable state reset within the execution time of half of a stabiliser check, and thus minimise any charge noise injection into the mediator and rest of the circuit.   

Without the use of mediators, leakage errors apply directly to the qubit dots and require leakage correction schemes to be applied. As discussed in Appendix~\ref{sect:leakage_comparison}, such schemes would introduce large qubit/runtime overheads~\cite{aliferisFaulttolerantQuantumComputation2007, sucharaLeakageSuppressionToric2015}, limits on the choice of data/ancilla qubits~\cite{fowlerCopingQubitLeakage2013} and/or require extra components for charge detection or reset introduced within a potentially dense qubit array. In contrast, in the architecture we propose here the leakage errors are addressed by the inherent charge deficit transfer from the qubits to the mediators and resetting the mediators using charge reservoirs. No additional components are needed since the reservoirs are also used for qubit initialisation and no additional runtime is introduced since the mediator resets can be carried out in parallel with other error checking cycles in the surface code.

\section{Surface code simulation}\label{sect:surface_code}
\subsection{Surface Code Threshold and Stabiliser Check Circuit}\label{sect:stb_circuits}
For many quantum error correction codes, there exists a threshold such that if the error rate of the physical circuit components falls within this threshold, then the logical error rate can be indefinitely reduced by scaling up the code size. Such an error threshold is highly dependent on the precise implementations of the quantum error correction circuits and the errors associated with their component parts. By transforming all the noise channels into Pauli channels via twirling~\cite{caiConstructingSmallerPauli2019}, we can efficiently simulate quantum error correction circuits using classical computers exploiting the Gottesman-Knill theorem~\cite{gottesmanHeisenbergRepresentationQuantum1998, aaronsonImprovedSimulationStabilizer2004} to obtain a reliable threshold for a given quantum error correction code~\cite{gellerEfficientErrorModels2013, gutierrezApproximationRealisticErrors2013, gutierrezComparisonQuantumErrorcorrection2015}. The threshold sets a target error rate for the experimentalist to aim for in order to implement a given quantum error correction code, though operation well below the threshold is required for useful quantum computing to be performed.

The surface code is implemented by checking the $X$/$Z$ parities of the data qubits spanned by each plaquette in Figure~\ref{Fig:layout_schematics}. These parities are the stabiliser generators of the surface code and are measured using the stabiliser-check circuits. Surface codes under depolarising gate noise using various stabiliser-check circuits can have a threshold in the range of $0.5\%$ -- $1\%$~\cite{stephensFaulttolerantThresholdsQuantum2014}.
\begin{figure}[t]
    \centering
    \begin{tikzcd}
        \lstick[wires=2]{\text{data}}&\lstick{1}&\gate{Y_{-\frac{\pi}{2}}}\gategroup[wires=2,steps=1,style={dashed, inner xsep=4pt, inner ysep=4pt}]{}&\ctrl{2} &\qw      &\gate{Y_{\frac{\pi}{2}}}\gategroup[wires=2,steps=1,style={dashed, inner ysep=4pt, inner xsep=-6pt}]{}&\qw  \\
        &\lstick{2}&\gate{Y_{-\frac{\pi}{2}}}&\qw     &\ctrl{1}      &\gate{Y_{\frac{\pi}{2}}}&\qw\\
        &\lstick[wires=2]{\text{anc: }$\ket{S}$}&\lstick{1}&\ctrl{}&\ctrl{}&\gate[wires=2, style={inner ysep=0pt, yshift=-4pt}]{\parbox{4em}{S-T\\Readout}}\\
        &&\lstick{2}&\ctrl{}&\ctrl{}& \\
        \lstick[wires=2]{\text{data}}&\lstick{3}&\gate{Y_{-\frac{\pi}{2}}}\gategroup[wires=2,steps=1,style={dashed, inner xsep=4pt, inner ysep=4pt}]{}      &\ctrl{-1}         &\qw      &\gate{Y_{\frac{\pi}{2}}}\gategroup[wires=2,steps=1,style={dashed, inner ysep=4pt, inner xsep=-6pt}]{}&\qw\\
        &\lstick{4}&\gate{Y_{-\frac{\pi}{2}}}      &\qw	          &\ctrl{-2} &\gate{Y_{\frac{\pi}{2}}}&\qw
    \end{tikzcd}
    \caption{{\bf The stabiliser check circuit}. The $Y$ rotations in the dashed boxes are only included during $X$ stabiliser checks. The two ancillae are initialised in a singlet state $\ket{S}$, and measured using Pauli spin blockade (singlet/triplet dependent tunnelling).}
    \label{Fig:stb_circuit}
\end{figure}
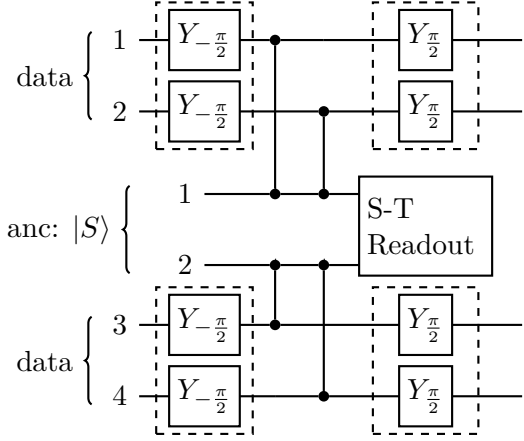

Our stabiliser-check circuit is shown in Figure~\ref{Fig:stb_circuit}, where the CZ gates must be further decomposed into $\sqrt{\text{SWAP}}$ or $S$ as outlined in Section~\ref{sect:two_qubit_gate}.
Besides $\sqrt{\text{SWAP}}$ or $S$, we also need single-qubit $Z$ rotations to construct CZ. $Z$ rotations can be implemented as a combination of $X$ and $Y$ rotations (which can be slow as noted in Section~\ref{sect:micromagnet}), or using the Stark shift whose speed is limited by the detuning range and whose accuracy relies on careful calibration. Fortunately, in our stabiliser-check circuit, most of the $Z$ rotations on the data qubits can be implemented in a virtual way by shifting the phases of all the future single-qubit rotations pulses~\cite{mckayEfficientZGatesQuantum2017}, and the $Z$ rotations on the ancillae can be omitted since we are performing symmetric operations on the singlet subspace (see Appendix~\ref{sect:virtual_Z} for details). The only $Z$ rotation that we need to explicitly implement is the $Z_\pi$ sandwiched by the two $\sqrt{\text{SWAP}}$s, applied to the 
data qubits.
This optimisation to remove single qubit gates substantially reduces the runtime and depth of the stabiliser-check circuit.

As shown in Figure~\ref{Fig:stb_circuit}, our circuit applies $\sqrt{Y}$ to all data qubits to switch between $X$ and $Z$ stabiliser checks. Because the ancillae are initialised in singlet states, the $\sqrt{Y}$ operations can be achieved using global ESR operations applied to all spins~\ref{sect:anc}, or through local operations applied only to the data qubits, using EDSR. 

\subsection{Stabiliser Cycle and Error Model}\label{sect:stb_cycle}
\begin{figure}[t]
    \centering
    \includegraphics[scale = 0.6]{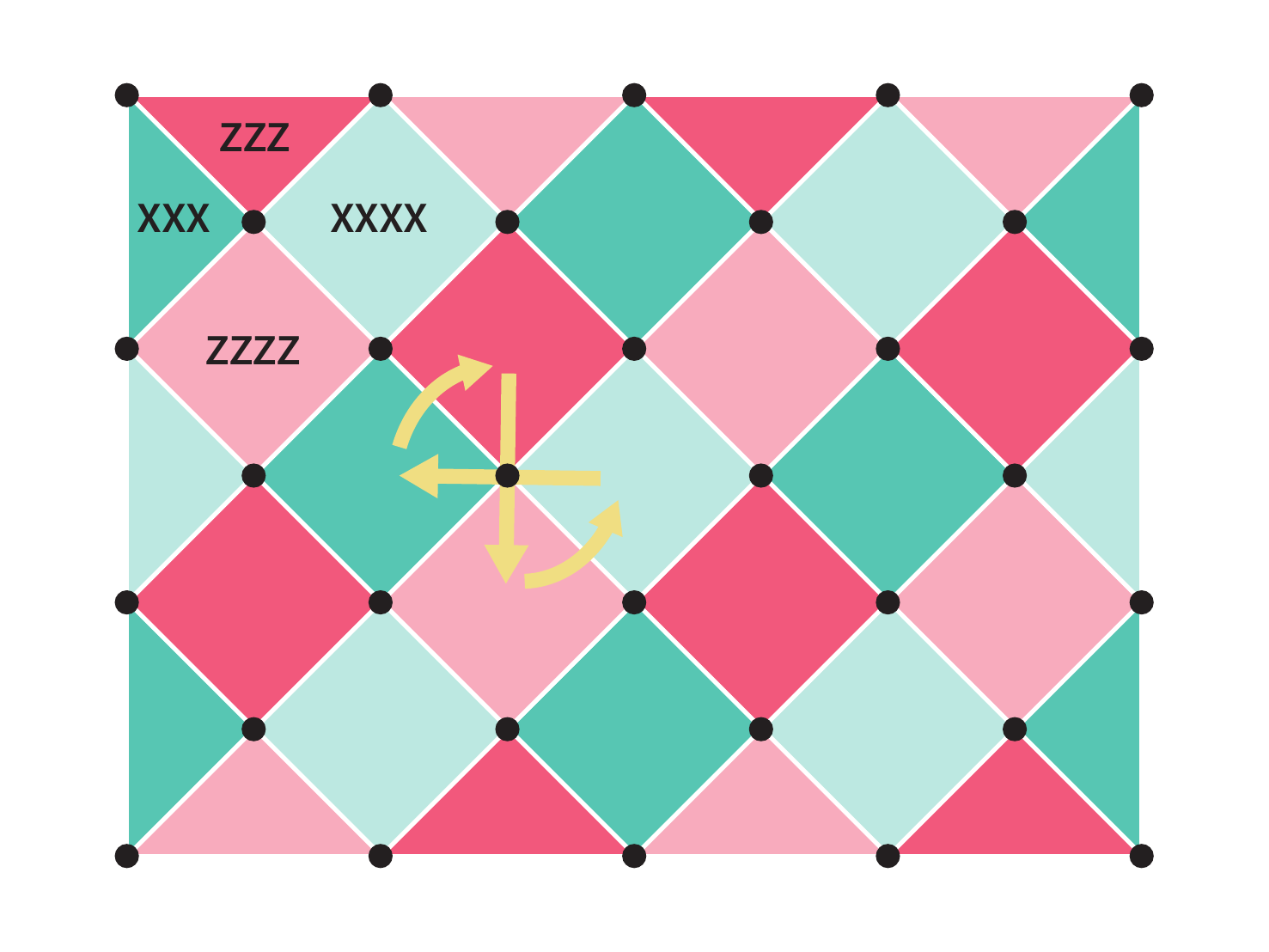}
    \caption{{\bf Ordering of stabiliser check cycles.} Each plaquette is given one of four colours, such that plaquettes of the same colour share no data qubits between them. Stabiliser checks of all plaquettes of a given colour are carried out simultaneously, in the sequence indicated by the arrows.}
    \label{Fig:stb_partitions}
\end{figure}
We divide all stabiliser checks into four disjoint partitions, performed in sequence, as shown in Figure~\ref{Fig:stb_partitions}. When one of the partitions become active, any two different stabiliser checks within it are separated by at least one inactive plaquette, across which leakage error cannot propagate. Hence, within each partition, errors (including  leakage errors) of one stabiliser check are independent of that of another stabiliser check, such that there are no spatial error correlations beyond a given plaquette.
During the stabiliser check of one partition, the mediator reset operation can be activated in the other partitions (see Section \ref{sect:robust_against_charge_leakage}). In this way, leakage errors arising during the active cycle of a given  partition do not survive to its subsequent cycle, removing the potential for temporal error correlations.
Using this partitioning and sequence of stabiliser updates, the errors in each stabiliser check should be Markovian, removing the temporal and spatial correlations in noise that can be highly damaging to the surface code, and greatly simplifying our error simulation.

Within each stabiliser check, we assume the following error model:
\begin{itemize}
    \item \textbf{Two-qubit gates:} Charge noise leads to fluctuations in the exchange strength $J$, which can lead to the following errors for the two-qubit gates that we are considering (shown in Appendix~\ref{app:two_qubit_err}):
    \begin{itemize}
        \item 
        $S$ gate has $Z_1Z_2$ error with probability $p_{2}$.
        \item 
        $\sqrt{\text{SWAP}}$ gate has SWAP error with  probability $p_{2}/2$. 
    \end{itemize} 
    To allow a simple comparison of the thresholds of the two kinds of two-qubit gates, we formulate our simulations in terms of a two-qubit gate error rate $p_2$ equal to that of the $S$ gate. Assuming that $S$ gates take twice the time required by $\sqrt{\text{SWAP}}$, the variance of the exchange phase $Jt$ accumulated due to fluctuations in $S$ is twice  that of $\sqrt{\text{SWAP}}$, and thus the error probability of $S$ is twice of that of $\sqrt{\text{SWAP}}$ (see Appendix~\ref{app:gate_error_comparisons}).
    
    \item \textbf{Readout:} The current state-of-the-art $\mu$s-scale readout scheme can achieve $98\%$ fidelity~\cite{zhengRapidGatebasedSpin2019}, which is the same as the best two-qubit gate fidelity achieved~\cite{huangFidelityBenchmarksTwoqubit2019}. Hence, here we will assume the readout error rate can be improved at the same pace as two-qubit gate error rate so that we have $p_{readout} = p_2$.

    \item \textbf{One-qubit gates and initialisation} are assumed to have a common depolarising error probability $p_{1}$. 
    The fidelity of one-qubit gates is typically more than one order of magnitude better than two-qubit gates~\cite{veldhorstAddressableQuantumDot2014, veldhorstTwoqubitLogicGate2015, watsonProgrammableTwoqubitQuantum2018}, thus we assume $\frac{p_1}{p_2} = 0.1$. 
    
    \item \textbf{Spin Leakage:} 
    As mentioned in Section~\ref{sect:robust_against_spin_leakage}, spin leakage in the ancilla qubits can be taken into account by considering all the possible errors on the individual spin within the ancilla spin pairs. Its effect on the measured parity can also be considered by flipping the parity result whenever there are asymmetric noise acting on the ancilla spin pairs. Note that is a more damaging noise model than the rigorous model\footnote{E.g. if the correct state before the readout is the spin-zero triplet state, then even if leakage errors take our state into other triplet state, our readout result should still be correct even though a leakage due to asymmetric noise has happened}, thus should give us a lower bound on the threshold.
    
    \item \textbf{Charge Leakage:} 
    When considering charge leakage errors, we first note that each stabiliser check can be divided into two non-interacting halves, each with one ancilla dot interacting with two data dots via two mediator dots as shown in Figure~\ref{Fig:layout_schematics}. Within each half of the stabiliser, when a leakage error occurs and get restored by the mediators, we will assume the worst-case left-over computational errors in which we have depolarising errors on the whole half (on both of the data qubit and the ancilla spin), so that the leakage error thresholds we derive below can be taken as a lower bound. 
    
    Charge leakage errors are most likely to occur during the tuning of potentials, thus we will assume here the charge leakages will only occur during the CZ gates in the stabiliser checks. If $p_{\rm leak}$ is the probability that a charge leakage error occurs during a CZ gate, then needing to perform two CZ gates in each half of the stabiliser checks means that there is a $2p_{\rm leak}$ probability that the whole half of the stabiliser check will get depolarised in each round of error check.
    
\end{itemize}

\begin{figure*}[t]
    \centering
    \subfloat[]{\includegraphics[width = 0.5\textwidth]{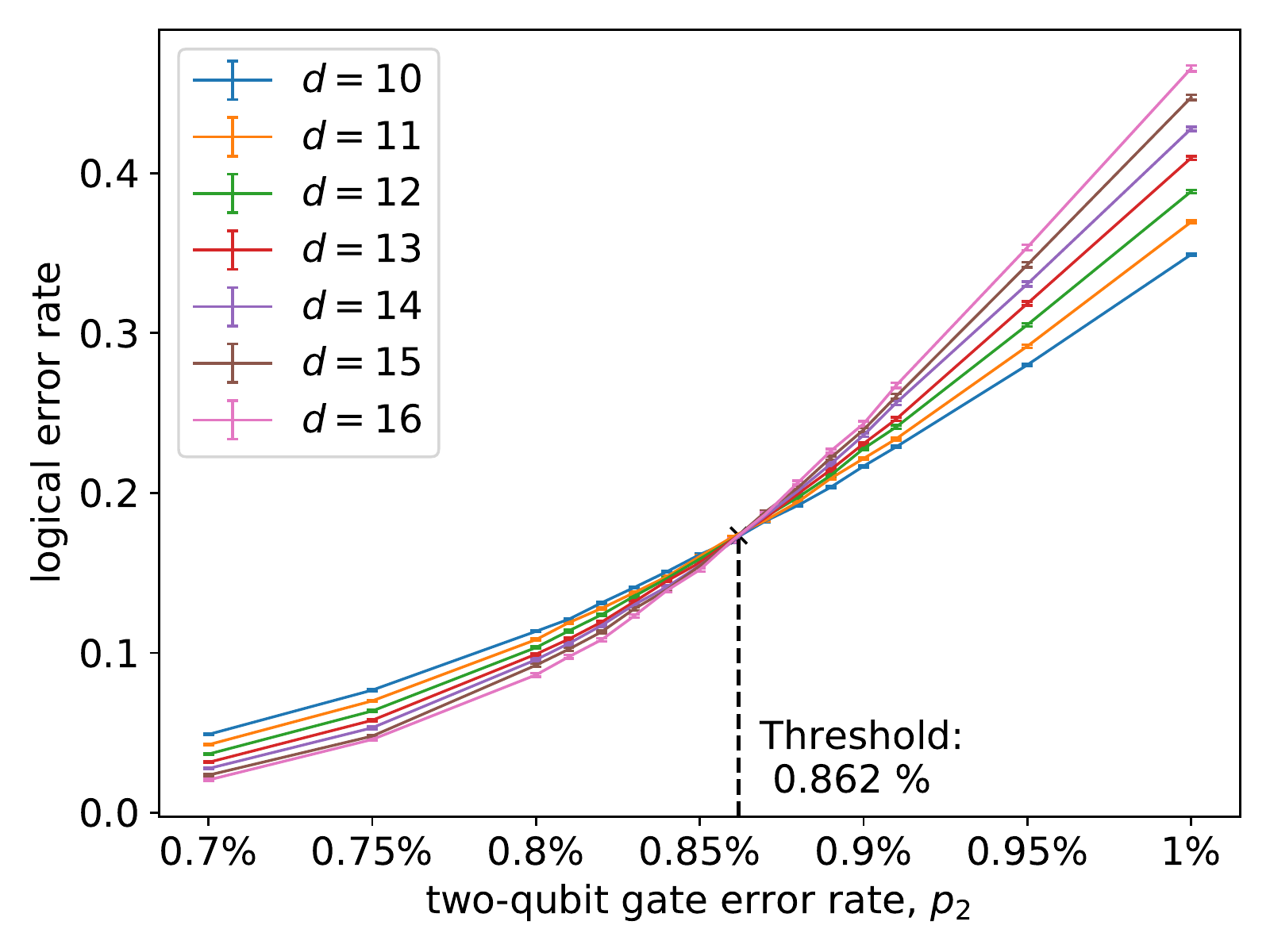}}
    \subfloat[]{\includegraphics[width = 0.5\textwidth]{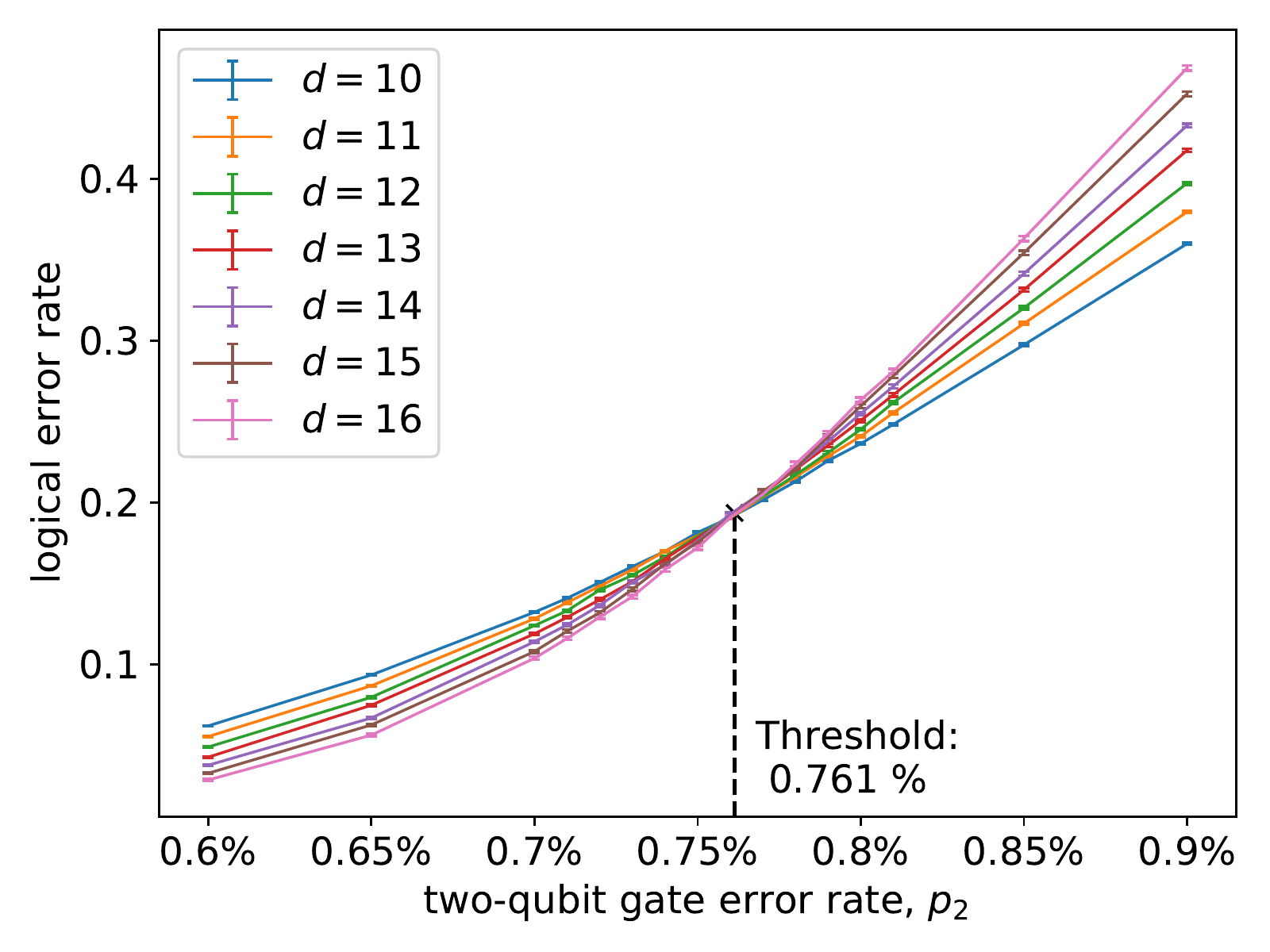}}\\
    \caption{Surface code two-qubit gate error threshold calculations in the case of no leakage error ($p_{\rm leak} = 0$) assuming (a) $S$ gates with error rate $p_2$ or (b) $\sqrt{\text{SWAP}}$ gates with error rate $p_2/2$.
        In all calculations, the error rate of single-qubit gates ($p_1$) and two qubit gates ($p_2$) is assumed to be fixed $\left(\frac{p_1}{p_2} = 0.1\right)$. $d$ is the code distance of the surface code.}
    \label{Fig:threshold_plot_two}
\end{figure*}

\subsection{Surface Code Threshold Results}\label{sect:threshold_result}
First we consider cases without charge leakage errors ($p_{\rm leak} = 0$). As shown in Figure~\ref{Fig:threshold_plot_two} (a) and (b), the threshold for $p_2$ is $0.86\%$ using $S$ gates and $ 0.76\%$ using $\sqrt{\text{SWAP}}$. Both are comparable to the threshold $ 0.75\%$ obtained using simple depolarising noise model~\cite{raussendorfTopologicalFaulttoleranceCluster2007}. The lower threshold for the architecture using $\sqrt{\text{SWAP}}$ is primarily due to the additional $Z_\pi$ needed to construct the CZ gate. Note that the gate errors here also include the spin leakage errors of the ancillae.

\begin{figure*}[t]
    \centering
    \subfloat[]{\includegraphics[width = 0.5\textwidth]{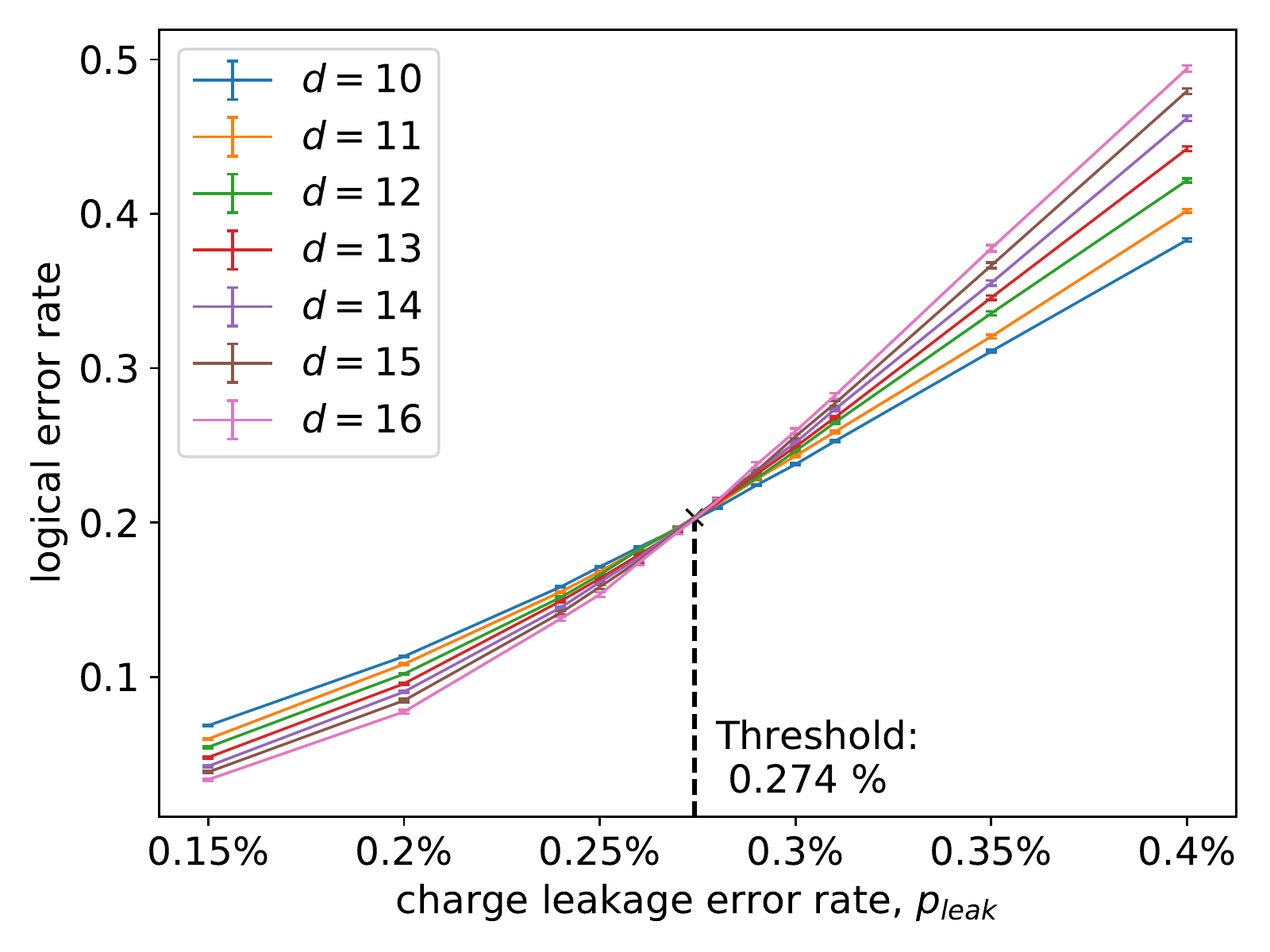}}
    \subfloat[]{\includegraphics[width = 0.5\textwidth]{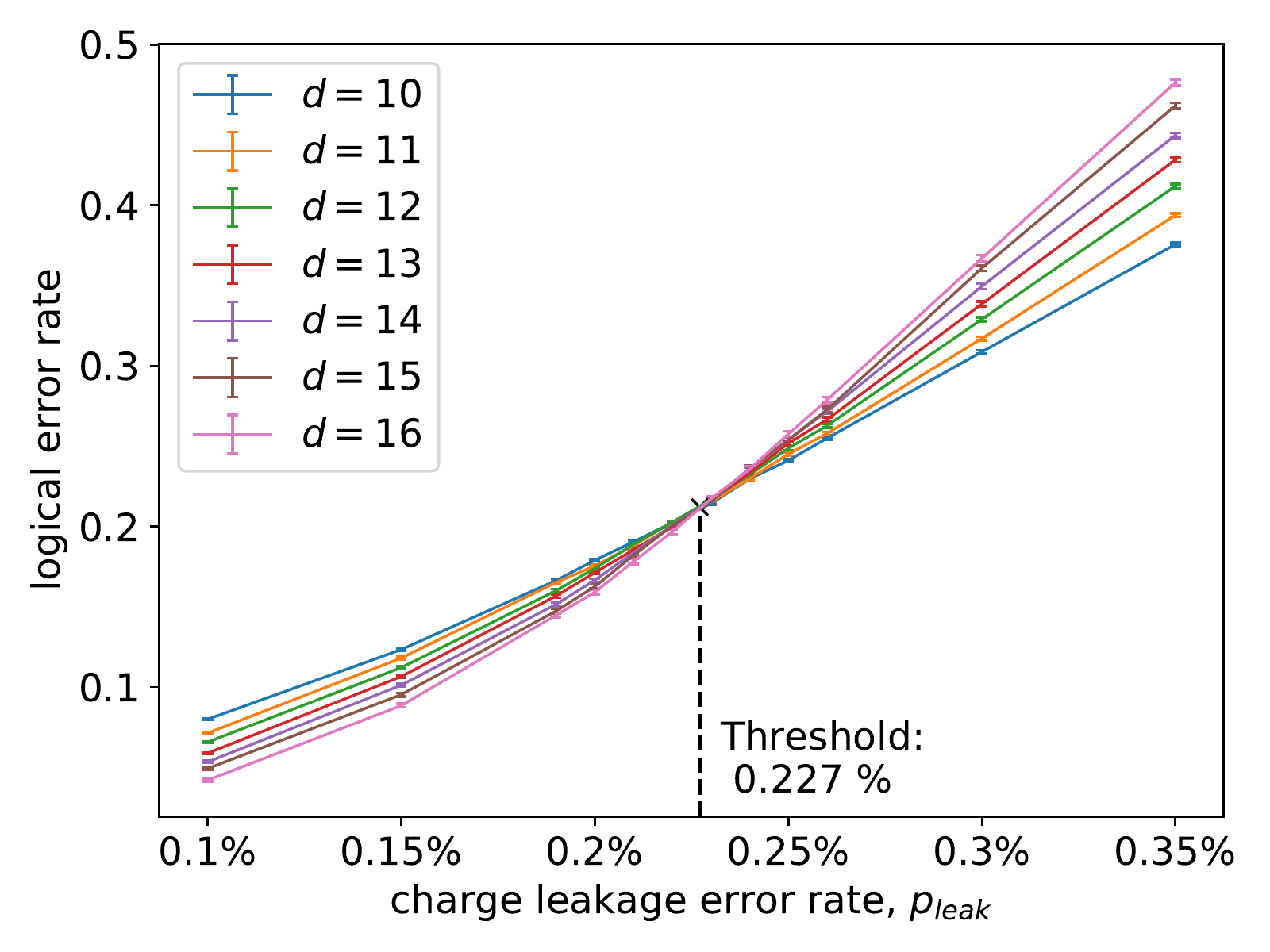}}\\
    \subfloat[]{\includegraphics[width = 0.5\textwidth]{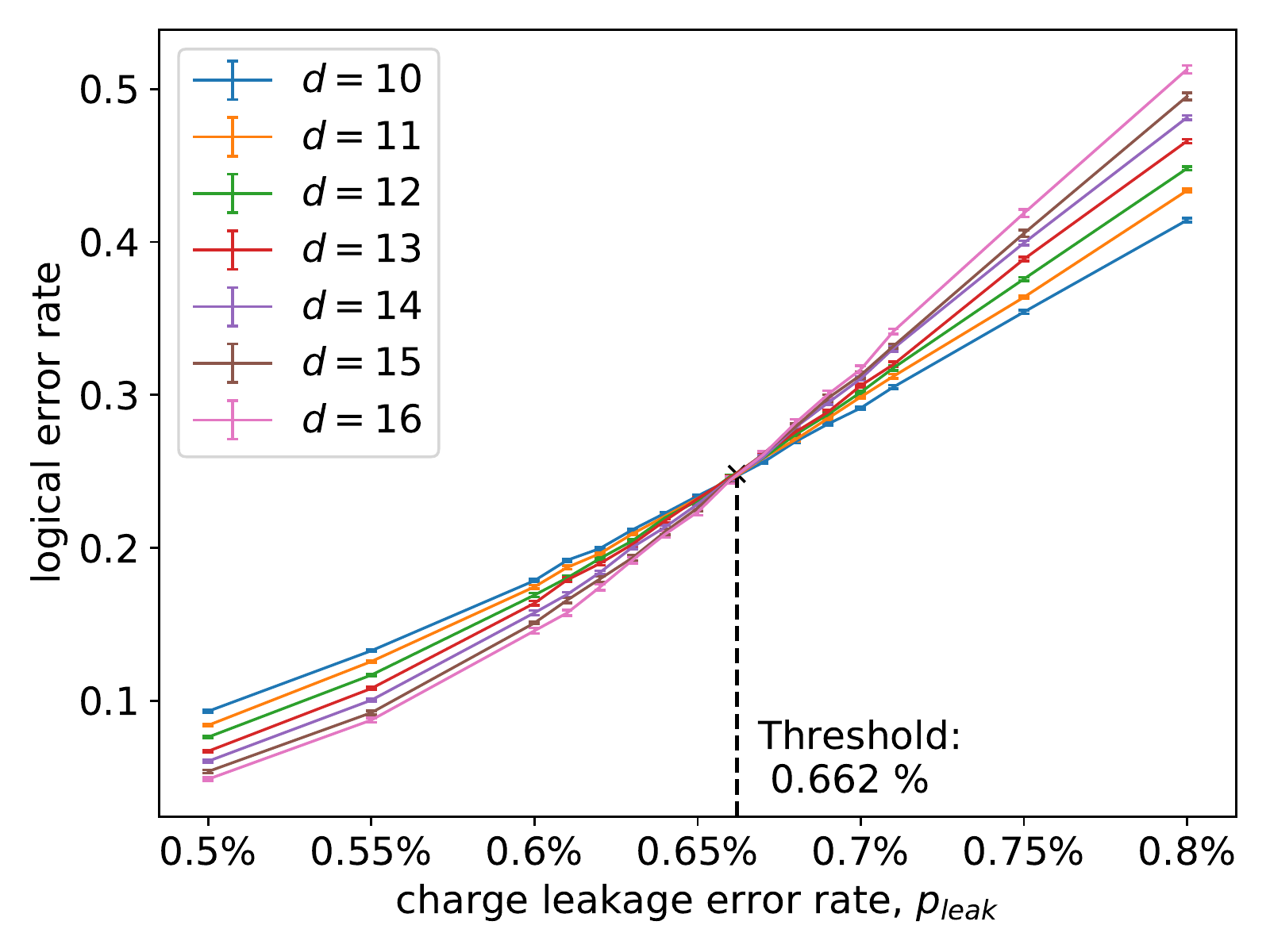}}
    \caption{Surface code leakage error $p_{\rm leak}$ threshold calculations assuming the use of (a) $S$-gates with error probability $p_{2} = 0.5\%$, (b) $\sqrt{\text{SWAP}}$ gates with error rate $p_2/2=0.25\%$, or (c) perfect gates ($p_2=0$).
        In all calculations, the error rate of single-qubit gates ($p_1$) and two-qubit gates ($p_2$) is assumed to be fixed $\left(\frac{p_1}{p_2} = 0.1\right)$. $d$ is the code distance of the surface code.}
    \label{Fig:threshold_plot_leak}
\end{figure*}

To achieve fault-tolerant quantum computation, our gate error rate need to be below the gate error thresholds. Suppose we manage to achieve a gate error rate below these thresholds at $p_2 = 0.5\%$, then the level of charge leakage error we can tolerate with such a gate error rate are indicated by the $p_{\rm leak}$ threshold in Figure~\ref{Fig:threshold_plot_leak} (a) and (b), which are $0.27\%$ with $S$ gates and $0.23\%$ with $\sqrt{\text{SWAP}}$. The charge leakage thresholds we obtained here are on the same order as the gate error rate we assumed here. The energy barrier of the charge leakage errors is usually higher than the errors in the spin space. Hence, we will expect the charge leakage error rate to be much lower than the usual gate error rate and thus below the charge leakage thresholds we obtained here.

If we can further push down the gate error rate (reducing $p_2$), the charge leakage error threshold will grow, and in the end bounded by the limit in the case of no gate errors ($p_2 = 0$) where the threshold for $p_{\rm leak}$ is $ 0.66\%$  (see  Figure~\ref{Fig:threshold_plot_leak} (c)). The similarity of this pure charge leakage error threshold to that from depolarising noise threshold indicates that in our architecture charge leakage errors can be effectively reduced to computational errors (i.e.\ errors within the computational subspace) via charge relaxation. In other words, even though the resultant computational errors have strong correlations within a given stabiliser check, charge leakage errors can be limited to be no more damaging than other conventional gate errors. The trade-off between the charge leakage threshold and the gate error rates is further illustrated by additional threshold simulations in Appendix~\ref{sect:further_sim}. Overall, this architecture shows good tolerance towards the computational errors resulting from charge leakage errors, even with a reasonable amount of gate errors present.

\subsection{Decoherence Errors}
We denote the characteristic time scale of the exchange interaction $T_J = \frac{\pi}{J}$, that of a Hadamard gate ($\sqrt{Y}$) as $T_H$ and that of a Z gate as $T_Z$. Using the stabiliser cycle outlined in Section~\ref{sect:stb_cycle} and the stabiliser circuit shown in Figure~\ref{Fig:stb_circuit}, the time needed for one stabiliser cycle is $T_S^{\rm cycle} = 8T_J + 2T_H$ assuming the use of $S$-gates and  $T_{\sqrt{\text{SW}}}^{\rm cycle} = 8T_J + 8T_{Z} + 2T_H$ with $\sqrt{\text{SWAP}}$. We have not accounted for the time required for initialisation and readout of the ancillae and the mediator resets because they can take place in parallel with other operations of the stabiliser circle. Such operations only become significant to the rate of the stabiliser check once they become an order of magnitude slower than the quantum gates, which is not the case in the range of parameters that we are considering (see Section~\ref{sect:anc} and \ref{sect:charge_reservoirs}). 

Using parameters outlined in Section~\ref{sect:two_qubit_gate}, we expect $T_J \sim$ 1 \textmu s. Based on demonstrated electrical tuning of the $g$-factor, we estimate the duration of Z gates implemented using Stark shifts to be $T_Z \sim$ 0.25 \textmu s~\cite{hwangImpactFactorsValleys2017}, while the time needed for a Hadamard gate is likely to differ depending on the use of ESR ($T_H \sim$ 1\textmu s) or EDSR ($T_H <$ 0.1\textmu s). We can therefore consider two illustrative cases for the stabiliser cycle time. In one case, micromagnets are used to enable the use of $S$-gates and EDSR, giving $T_{\rm fast}^{\rm cycle} \sim$ 8~\textmu s (limited by $T_J$). In the other limit, the slower ESR gates and less efficient $\sqrt{\text{SWAP}}$ are used, giving $T_{\rm slow}^{\rm cycle} \sim$ 12~\textmu s (limited by $T_H$, $T_Z$ and $T_J$).

For electron spins in quantum dots in isotopically-enriched silicon, decoherence times have been reported ranging from $T_2^* =$ 20~\textmu s and $T_{\rm 2,CPMG} =$ 3~ms in systems with a  micromagnet~\cite{yonedaQuantumdotSpinQubit2018} to $T_2^* =$ 120~\textmu s and $T_{\rm 2,CPMG} =$ 28~ms in systems without micromagnets~\cite{veldhorstTwoqubitLogicGate2015}. The probability of phase flip error \emph{per stabiliser cycle} using Carr-Purcell-Meiboom-Gill (CPMG) decoupling is hence  $\frac{T^{cycle}}{2T_2} \approx 2 \times 10^{-4}$ to $10^{-3}$ for the parameters we considered, well within the \emph{per gate} error threshold we obtained in Section~\ref{sect:threshold_result}. We conclude that the finite decoherence time of spins in silicon measured in devices to date can be tolerated by our surface code architecture.

\section{Conclusions and Outlook}
We have introduced a surface code architecture implemented using spin qubits in silicon quantum dots that is robust against spin leakage errors through its use of single-dot data qubit and robust against charge leakage errors through its use of multi-electron mediator dots.  Our approach efficiently unifies the task of maintaining a proper charge distribution (essential for any SS quantum device) together with the task of performing the stabiliser cycles required by the surface code.  Charge leakage from the qubit dots is transferred to the mediator dots via fast charge relaxation, and removed using charge reservoirs attached to the mediators, reducing the charge leakage errors to the level of standard computational errors that can be corrected by the surface code. 
We find that our stabiliser check cycle removes time and space correlations in the remaining computational errors, which can be highly damaging to surface codes. The depth of the stabiliser-check circuit was reduced by the symmetry of the double-dot ancillae and virtual Z gates. Through simulations, we find that the surface code threshold for the computational errors arising from charge leakage errors is $0.66 \%$ in the absence of gate errors, showing that its effect can be limited to that of standard depolarising gate errors. 
Under a reasonable gate error rate $0.5\%$ (which includes ancilla spin leakage errors), we obtain a charge leakage error threshold of $0.23 \sim 0.27\%$, showing good tolerance of our architecture towards charge leakage errors even under gate noise. 
The fidelity of two-qubit gates is expected to the principal bottleneck for reaching the fault-tolerant level, and experimentally demonstrating a high-fidelity mediated exchange interaction using isotopically enriched silicon will be a key step in validating this architecture.

Besides adding tolerance towards leakage errors, the elongated mediator dots in our structure also relax the density of the qubit dots, offering more space in-plane for the essential measuring devices, charge reservoirs and classical control lines, and facilitating fabrication using (e.g.) CMOS technology~\cite{vandersypenInterfacingSpinQubits2017}. 
The extra space provided by the mediators and the unique properties of the double-dot ancilla enable more convenient integration of micromagnets which can increase the speed of both the single qubit rotations and stabliser check cycle.

We find that gate mechanisms and energy scales that have already been experimentally reported will suffice to realise a stabiliser cycle time approaching the MHz domain, and that this speed is sufficient to suppress environmental decoherence. This is not a fundamental limit to the operation speed of such a device, however, it makes use of a mediated exchange interaction, which is inherently slower than direct exchange. To push the speed further, data or ancilla spins could be shuttled onto the mediators for direct exchange with a neighbouring ancilla/data spins, and such exchange gates between a single-electron dot and a multi-electron dot have been demonstrated~\cite{malinowskiFastSpinExchange2019, malinowskiSpinMultielectronQuantum2018}. Shuttling in combination with micromagnet-induced field gradients may introduce significant dephasing noise which may be challenging to correct (e.g.\ using calibration and single-qubit rotations). As in many approaches, there is a trade-off between speed and error rate to be carefully considered. 

A second factor in the speed of the processor operation is charge relaxation, which we have assumed to be fast compared to the gates. If this were not the case charge leakage errors would not be rapidly transferred from the qubit dots to the mediators, and empty qubit dots may remain after a stabiliser cycle, leading to a non-trivial errors of a non-Markovian nature. Nevertheless, we would expect the charge leakage process and the relaxation restoring force to  reach some equilibrium, leaving the proportion of the empty qubit dots in the surface code fixed. Further work could study the non-Markovian effects of the empty qubit dots, and the equilibrium value of the leaked qubit fractions under different charge leakage and relaxation models. Nevertheless, in cases where charge relaxation time was non-negligible, existing leakage correction protocols like active leakage detection and correction, or leakage reduction units could be adopted. Indeed, combining active methods with inherent robustness to leakage errors may be advantageous, especially if the native leakage rate is high.

Features from other silicon quantum computing architectures like shared control lines~\cite{vandersypenInterfacingSpinQubits2017, liCrossbarNetworkSilicon2018} and modularity~\cite{buonacorsiNetworkArchitectureTopological2019} could also be adopted into our structure, if challenges around the inhomogeneity of quantum dots and shuttling noise can be minimised. Conversely, the introduction of additional quantum dots and electrons into a system to create accessible metastable charge states could be adopted in other approaches to offer robustness against charge leakage errors.

\section*{Acknowledgements}
This work has been supported by Quantum Motion Technologies Ltd. SS and SF acknowledge the Engineering and Physical Sciences Research Council
(EPSRC) through the Centre for Doctoral Training in
Delivering Quantum Technologies (EP/L015242/1) and 
SCB acknowledges support from ESPRC grant EP/M013243/1 (the NQIT Quantum Hub).

\newpage
\appendix
\section{Two ways to achieve CZ between data and ancilla qubits} \label{app:CZ}

\subsection{Hamiltonian}
\noindent The two-spin Hamiltonian is:
\begin{align}\label{eqn:hamitonian}
{H} &= \underbrace{\frac{1}{2} \left(E_1Z_1 + E_2 Z_2\right)}_{\text{\parbox{8em}{\centering${H}_0$: Zeeman splitting}}} + \underbrace{\frac{J}{2} \text{SWAP}}_{\text{\parbox{7em}{\centering ${H}_{ex}$: exchange\\ interactions}}}
\end{align}
\noindent The Zeeman splitting $H_0$ can be further split into:
\begin{align*}
\underbrace{\frac{1}{2} \left(E_1Z_1 + E_2 Z_2\right)}_{\text{\parbox{8em}{\centering${H}_0$: Zeeman splitting}}}  &= \underbrace{\frac{E_z}{2} \left(Z_1 + Z_2\right)}_{\text{\parbox{8em}{\centering${H}_Z$: average Zeeman splitting}}} + \underbrace{\frac{\Omega}{2} \left(Z_1 - Z_2\right)}_{\text{\parbox{8em}{\centering${H}_\Delta$: Zeeman splitting gradient}}}
\end{align*}
where $E_z = \frac{E_1 + E_2}{2}$, $\Omega = \frac{E_1 - E_2}{2}$.

\subsection{$\Omega \ll J$: simple exchange interaction}\label{sect:sqrtswap}
Since $\Omega \ll J$, and $\left[H_{ex}, H_Z\right]  = \left[\text{SWAP}, Z_1 + Z_2\right] = 0$, 
\begin{align*}
H_{ex, I} = e^{iH_0t}H_{ex}e^{-iH_0t} = H_{ex}
\end{align*}
i.e. to perform the exchange interaction in the rotating frame is just the same as performing the exchange interaction in the lab frame.

The evolution operator due to ${H}_{ex}$ is given by:
\begin{align*}
{U}_{ex}(t)& = e^{-i{H}_{ex} t}  = e^{-i\text{SWAP} \frac{Jt}{2}}
\end{align*}
A SWAP gate corresponds to $\frac{Jt}{2} = \frac{\pi}{2}$, and a $\sqrt{\text{SWAP}}$ gate corresponds to $\frac{Jt}{2} = \frac{\pi}{4}$.

The error in applying the exchange interaction arising from imprecise pulse timing or charge fluctuations is analysed in Appendix~\ref{app:two_qubit_err}.

A CZ can be implemented using $\sqrt{\text{SWAP}}$ in the following way:
\begin{center}
    \begin{tikzcd}
        &\ctrl{1} &\qw \\
        &\ctrl{-1}&\qw
    \end{tikzcd}
    \begin{tikzcd}
        \\
        \equiv\\
    \end{tikzcd}
    \begin{tikzcd}
        &\gate{Z_{\frac{\pi}{2}}} &\gate[wires=2, style={yshift=-9.5pt, inner ysep=-5pt}]{\rotatebox{270}{$\sqrt{\text{SWAP}}$}}    &\gate{Z_\pi} &\gate[wires=2, style={yshift=-9.5pt, inner ysep=-5pt}]{\rotatebox{270}{$\sqrt{\text{SWAP}}$}}         &\qw\\
        &\gate{Z_{-\frac{\pi}{2}}} &     &\qw&  &\qw\\
    \end{tikzcd}
\end{center}

\subsection{$\Omega \gg J$: dipole-dipole interaction}\label{app:achieving_one_step_CZ_1}
Following arguments from~\cite{meunierEfficientControlledphaseGate2011, watsonProgrammableTwoqubitQuantum2018}, without exchange interaction we have
\begin{align*}
H_ 0 &= \frac{1}{2}\begin{pmatrix}
E_z&0&0&0\\0&\Omega &0&0\\0&0&-\Omega&0\\0&0&0&- E_z
\end{pmatrix}
\end{align*}
We can see that $E_z$ determine the eigenenergies in the parallel spin subspace, while $\Omega$ determine the eigenenergies in the anti-parallel spin subspace.

If we add in the exchange Hamiltonian
\begin{align}
{H}_{ex} & = \frac{J}{2}\begin{pmatrix}
1&0&0&0\\0&0 &1&0\\0&1&0&0\\0&0&0&1
\end{pmatrix},
\end{align}
in the parallel spin subspace, the energy of both states will be shifted up by $\frac{J}{2}$. In the anti-parallel spin subspace, if $\Omega \gg J$, then $H_{ex}$ can be treated as perturbation. Using first order perturbation theory, the shift in eigenenergies for the anti-parallel spin states is 0.

Hence, to first order approximation, in which the eigenstate do not change and only eigenenergies change, the exchange Hamiltonian (which is to first order the shift in eigenenergies) becomes
\begin{align}
{H}_{ex} & = \frac{J}{2}\begin{pmatrix}
1&0&0&0\\0&0 &0&0\\0&0&0&0\\0&0&0&1
\end{pmatrix}.
\end{align}
This is just a \emph{dipole-dipole} interaction, which, because it commutes with $H_0$, has a rotating frame form identical to its lab form.

Allowing this Hamiltonian to evolve for a time period $\frac{\pi}{J}$, produces the following gate:
\begin{align*}
S  \propto \begin{pmatrix}
1 &0 &0 &0\\
0 &-i &0 &0\\
0 &0 &-i &0\\
0 &0 &0 &1
\end{pmatrix}
\end{align*}
A CZ gate can be built from  $S$ using:
\begin{center}
    \begin{tikzcd}
        &\ctrl{1} &\qw \\
        &&\\
        &\ctrl{-1}&\qw
    \end{tikzcd}
    \begin{tikzcd}
        \\
        \equiv\\
    \end{tikzcd}
    \begin{tikzcd}
        &\gate{Z_{\frac{\pi}{2}}}&\gate[wires=2]{S} &\qw       \\
        &\gate{Z_{\frac{\pi}{2}}}& &\qw 
    \end{tikzcd}
\end{center}

\subsection{Virtual Z gate and symmetric operations on ancilla}\label{sect:virtual_Z}
Whether using $S$ or $\sqrt{\text{SWAP}}$ to construct a CZ, the only type of single-qubit gate needed is the $Z$ rotation, which can be implemented in a virtual way by shifting the rotating reference frame by a given phase~\cite{mckayEfficientZGatesQuantum2017}. Such $Z$ rotations are essentially error-free and require zero time. This corresponds to adding a phase offset to all subsequent $X$, $Y$ gate pulses, and switching all subsequent two-qubit gates into the new rotating frame after the virtual $Z$ rotation. Two-qubit gates whose Pauli components consist of only tensor products of $I$ and $Z$ are invariant under changing rotating reference frame, hence we do not need to modify these two-qubit gates after the virtual $Z$ rotation. The other two-qubit gates usually have different forms in the shifted rotating frame and might not be achievable through our Hamiltonian.

Following such arguments, we find that for the CZ gate constructed using the exchange-interaction, the $Z_\pi$ bracketed by the two $\sqrt{\text{SWAP}}$s cannot be applied in a virtual way, while the two $Z$ rotations outside the $\sqrt{\text{SWAP}}$s can. For the dipole-dipole CZ gate, all the $Z$ rotations can be applied in a virtual way.

However, there is another caveat. For the virtual $Z$ rotation to work, we need to do the measurements in $Z$ basis at the end, so that all the remnant $Z$ rotation for compensating for the virtual $Z$ gates will have no effect on the measurements (though we can use the shifted one qubit gate to change the measurement basis). Our ancilla measurement does not use a standard basis: our measurement only tells us whether the ancilla is in the singlet or triplet state, where the singlet state and the triplet states does not corresponds to a qubit representation. Thus, we cannot use virtual $Z$ gates here for our ancilla qubits, but can instead permute all the $Z$ rotations (besides the one bracketed by $\sqrt{\text{SWAP}}$) to the position right after the initialisation of the singlet state. We then use the fact that the initial singlet state is invariant under symmetric gates operating on both ancilla dots, to see that there is no need to apply the $Z$ rotations at the ancilla (besides the one bracketed by $\sqrt{\text{SWAP}}$). 

Hence, under either approach to implement a CZ gate, the only single-qubit gate that we need to implement is the $Z_\pi$ bracketed by $\sqrt{\text{SWAP}}$s. All the other $Z$ rotations can be either implemented in a virtual way or can be omitted due to the property of our ancilla qubits.

\subsection{Comparison of the two implementations of CZ}
\subsubsection{Operation time}

We denote the characteristic time scale of exchange interaction as $T_J = \frac{\pi}{J}$, and that of Z gate as $T_Z$. The time we needed to achieve a CZ using dipole-dipole like interaction is just $T_J$, no single-qubit gates needed. On the other hand, the time we need to achieve a CZ using exchange interaction is $T_J + T_Z$. The extra term here is due to the $Z_\pi$ gate that we need to explicitly implement.

\subsubsection{Errors}\label{app:gate_error_comparisons}
\noindent\textit{Errors due to fluctuation of $Jt$:}\\

The ideal exchange phase for $\sqrt{\text{SWAP}}$ is $\theta_{sw} = Jt_{sw} = \frac{\pi}{2}$. We will denote the variance in $\theta_{sw}$ due to fluctuations in exchange strength $J$ or operation time $t$ as $\epsilon_{sw}^2$.

The ideal exchange phase for $S$ is $\theta_{s} = Jt_{s} = \pi$. If we divide the accumulation of phase $\theta_{s}$ into two independent stages, with each stage accumulating phase $\frac{\pi}{2} = \theta_{sw}$, then we have $\theta_{s} = \theta_{sw, 1} + \theta_{sw, 2}$. Hence, the variance of $\theta_{s}$ is just $\epsilon_{s}^2 = 2\epsilon_{sw}^2$.

As shown in Appendix~\ref{app:two_qubit_err}, such fluctuations will lead to:
\begin{itemize}
    \item $\sqrt{\text{SWAP}}$: $p_{sw} = \epsilon_{sw}^2$ probability of having a swap error.
    \item $S$: $p_{s} = \epsilon_{s}^2 = 2p_{sw}$ probability of having a $Z_1Z_2$ error.
\end{itemize}

\noindent\textit{Errors due to approximations made:}\\

The main approximation made in deriving the exchange interaction is ignoring the higher order exchange terms which will not change the form of interaction (shift of energy in the singlet subspace w.r.t.\ the triplet subspace), but only shift the strength of exchange interaction. This is possible to overcome via careful calibrations. Of course there are also perturbations to the eigenstates that we have not considered, which might lead to leakage errors as shown in Appendix~\ref{app:leakage_mechanism}.

Since both $\sqrt{\text{SWAP}}$ and $S$ make use of exchange interactions, they are equally affected by the approximations made in the treatment of the exchange interaction. In addition, there are higher order corrections to the $S$ gate due to the assumption $J \ll \Omega$ of magnitude $\frac{J}{\Omega}$. Similarly, there are higher order corrections to the $\sqrt{\text{SWAP}}$ gate due to the assumption $\Omega \ll J$ of magnitude $\frac{\Omega}{J}$.

\section{Errors due to fluctuation in interaction strength and time}\label{app:two_qubit_err}
\subsection{General theory}
Suppose the Pauli basis of Hamiltonian $H$ is the set $G_H$:
\begin{align*}
H = \sum_{g_i \in G_H} \beta_ig_i
\end{align*}
note that $\beta_i$ are real since $H$ is Hermitian.

Then we can define the magnitude of $H$ to be $E$, and the normalised version of $H$ to be $h$ where:
\begin{align}
E &= \sqrt{\sum_i \beta_i^2}\\
h &= \frac{H}{E} = \sum_{{g}_i \in G_{{H}}} \frac{\beta_i}{E} {g}_i = \sum_{{g}_i \in G_{{H}}} \alpha_i {g}_i \label{eqn:h_decomp}
\end{align}
for $\alpha_i = \frac{\beta_i}{E}$ and we have $\sum_i \alpha_i^2 = 1$.

Now the evolution operator is just:
\begin{align*}
{U}(t)& = e^{-i{H} t}  = e^{-i{h} Et}\\
{U}(\theta) &= e^{-i\theta{h}}
\end{align*}
with $\theta = Et$. 

However, over- and under-rotations of $\theta$ occur in the experiment due to imprecise pulse timing $t$ or fluctuation of interaction strength $E$. If there is a $50\%$ percent chance of over and under rotation by $\epsilon\ll 1$, we have:
\begin{align*}
{U}(\theta \pm \epsilon) &= e^{-i \left(\theta \pm \epsilon\right) {h} }\\
& \approx e^{-i\theta{h} } \left({I} \mp i\epsilon{h} - \frac{\epsilon^2}{2} {h}^2\right)\\
& = {U}(\theta)\left({I} - \frac{\epsilon^2}{2} {h}^2 \mp i\epsilon{h} \right)
\end{align*}
Then the effective operation is just
\begin{align}
\mathcal{U}_{\theta, \epsilon}(\rho) &= \frac{1}{2}{U}(\theta + \epsilon) \rho {U}^\dagger(\theta + \epsilon) + \frac{1}{2} {U}(\theta - \epsilon) \rho {U}^\dagger(\theta - \epsilon)\nonumber\\
&=\left({I} - \frac{\epsilon^2}{2} {h}^2 \right){U}(\theta) \rho {U}^\dagger(\theta) \left({I} - \frac{\epsilon^2}{2} {h}^2 \right) \nonumber\\
&\qquad \qquad \qquad + \epsilon^2{h}{U}(\theta) \rho {U}^\dagger(\theta){h}\label{eqn:fluc_channel}
\end{align}
Similar channels are obtained for other symmetric over/under-rotation distributions that are centred on the correct rotation angles.
\subsubsection{${h}$ is unitary} \label{app: simple_swap_err}
If ${h}$ is unitary (and remember it is also Hermitian since it is the normalised Hamiltonian), e.g. $h$ is SWAP or Pauli, then (\ref{eqn:fluc_channel}) turns into 
\begin{align}
\mathcal{U}_{\theta, \epsilon}(\rho) 
& = \left(1 - \epsilon^2\right){U}(\theta) \rho {U}^\dagger(\theta)  + \epsilon^2{h}{U}(\theta) \rho {U}^\dagger(\theta){h} \label{eqn:fluc_err_uni}
\end{align}
i.e. we have either perfect ${U}(\theta)$ or $\epsilon^2$ probability of having a ${h}$ error on top of ${U}_{ex}(\theta)$.

\subsubsection{Twirling}
Twirling is a technique use for transforming the given error channel into a Pauli channel to obtain a simpler description of the error channel.

The Pauli decomposition of ${I} - \frac{\epsilon^2}{2} {h}^2 $ is
\begin{align*}
{I} - \frac{\epsilon^2}{2} {h}^2 &= {I} - \frac{\epsilon^2}{2} \left[\sum_{i,j} \alpha_i\alpha_j {g}_i{g}_j\right]\\
& = (1- \frac{\epsilon^2}{2}) {I} - \frac{\epsilon^2}{2} \left[\sum_{i\neq j} \alpha_i\alpha_j {g}_i{g}_j\right]
\end{align*}
After twirling, the noise due to non-identity Pauli components scales as $O(\epsilon^4)$ in the Pauli channel, and hence is negligible.

The Pauli decomposition of ${h}$ is just (\ref{eqn:h_decomp}). Hence, after twirling, the effective error channel we have is just:
\begin{align}
\mathcal{U}_{\theta, \epsilon}(\rho)  &= (1- \frac{\epsilon^2}{2}){U}(\theta) \rho {U}(\theta)^\dagger(1- \frac{\epsilon^2}{2})\nonumber \\
& \qquad +  \epsilon^2 \left[ \sum_{{g}_i \in G_{{H}}} \alpha_i^2 g_i {U}(\theta) \rho {U}(\theta)^\dagger g_i\right]\nonumber\\
& = (1- \epsilon^2){U}(\theta) \rho {U}(\theta)^\dagger\nonumber\\
& \qquad +  \epsilon^2 \left[ \sum_{{g}_i \in G_{{H}}} \alpha_i^2 g_i {U}(\theta) \rho {U}(\theta)^\dagger g_i\right] \label{eqn:fluc_err_non_uni}
\end{align}
i.e. it is an error channel with $\epsilon^2 \alpha_i^2 $ probability of the Pauli error $g_i$ happening on top of the perfect operation ${U}(\theta)$.

\subsection{Applications}
\subsubsection{Exchange Interaction}
For an exchange interaction, we have:
\begin{align*}
{H} = \frac{J}{2} \text{SWAP}
\end{align*}
We have fluctuation $\epsilon_{sw} \ll 1$ in $\theta = \frac{Jt}{2}$ and ${h} = \text{SWAP}$ is unitary. Hence, using (\ref{eqn:fluc_err_uni}), we have:
\begin{align*}
\mathcal{U}_{ex, \theta, \epsilon_{sw}}(\rho) & = \left(1 - \epsilon_{sw}^2\right){U}_{ex}(\theta) \rho {U}_{ex}^\dagger(\theta)  \\
&\qquad + \epsilon_{sw}^2 \text{SWAP}\ {U}_{ex}(\theta) \rho {U}_{ex}^\dagger(\theta)\ \text{SWAP}
\end{align*}
i.e. we have either perfect ${U}_{ex}(\theta)$ or $\epsilon_{sw}^2$ probability of having a $\text{SWAP}$ error on top of ${U}_{ex}(\theta)$.

\subsubsection{Dipole-dipole Interaction}
For a dipole-dipole interaction, we have:
\begin{align*}
{H} = \frac{J}{2} \left(Z_1Z_2\right)
\end{align*}
We have fluctuation $\epsilon_{s} \ll 1$ in $\theta = \frac{Jt}{2}$ and ${h} = Z_1Z_2$ is unitary. Hence, using (\ref{eqn:fluc_err_uni}), we have:
\begin{align*}
\mathcal{U}_{dd, \theta, \epsilon_{s}}(\rho) & = \left(1 - \epsilon_{s}^2\right){U}_{dd}(\theta) \rho {U}_{dd}^\dagger(\theta)  \\
&\qquad + \epsilon_{s}^2 Z_1Z_2\ {U}_{dd}(\theta) \rho {U}_{dd}^\dagger(\theta)\ Z_1Z_2
\end{align*}
i.e. we have $\epsilon_{s}^2$ probability of having a $Z_1Z_2$ error.

\section{Background exchange interaction}\label{app:background_ex}
In our system, $t_{ab}$ and $\Delta_M$ are generally fixed in a given device, however, their values can be engineered in the device design. The mediated exchange coupling (and hence the CZ gate) can be turned on and off by shifting the detuning of the mediator dot with respect to the side dots to switch $\Delta_{L/R}$ between $\Delta_{\rm on}$ and $\Delta_{\rm off}$. Since $\Delta_{\rm off}$ is finite, there is a residual exchange interaction even in the off stage. Using (\ref{eqn:exchange_strength}), we obtain the strength of such residual exchange interaction compared to our intended exchange interaction:
\begin{align*}
\frac{J_{\rm off}}{J_{\rm on}} = \left(\frac{\Delta_{\rm on}}{\Delta_{\rm off}}\right)^2
\end{align*}
If we look at the direct exchange interaction instead, we have $J \propto \frac{\abs{t}^2}{\Delta}$ and hence $\frac{J_{\rm off}}{J_{\rm on}} = \frac{\Delta_{\rm on}}{\Delta_{\rm off}}$. Hence, we see that the residual exchange interaction of mediated exchange is more suppressed than direct exchange when only tuning the on-site energy of quantum dots.

An imperfect `off' state also leads to next-nearest-neighbour interactions. For direct exchange interaction, the next nearest neighbour interaction is approximated as $\left(\frac{t}{\Delta_{\rm off}}\right)^2$ of the nearest neighbour interaction. In the mediated exchange interaction however, the next nearest neighbour interaction is approximately  $\left(\frac{t}{\Delta_{\rm off}}\right)^4$ of the nearest neighbour interaction, which is again much more heavily suppressed than the direct exchange case. 

Hence, by using mediated exchange interactions we can more confidently ignore the effect of residual exchange interactions and next nearest neighbour interactions in our analysis.

\section{Comparison of leakage resilience to architectures without mediators}\label{sect:leakage_comparison}
As mentioned before there are two general schemes to deal with leakage errors in qubits: using leakage reduction protocols or detecting leaked qubits and replacing them.

Using leakage reduction units~\cite{aliferisFaulttolerantQuantumComputation2007, sucharaLeakageSuppressionToric2015} requires a large number of additional ancilla qubits, which can be hard to integrate due to space constraints in addition to the qubit overhead they bring. We can reuse some of the ancilla qubits to alleviate such challenges, but this in turn significantly increases the surface code runtime and circuit depth. Another way to achieve leakage reduction is by swapping the data and ancilla qubits at the end of every full stabiliser cycle~\cite{fowlerCopingQubitLeakage2013}, which does not require any additional ancilla qubits. However, such a scheme is not compatible with architectures that have single-dot data qubit  and double-dot ancilla. Moreover, it assumes that the initialisation process of ancilla dots will fix the leakages. This will only be true if we use charge reservoirs for the initialisation of ancilla during the error correction cycle. To prevent the initialisation process of ancilla qubits from affecting other qubits, the charge reservoirs would need to be integrated into the structure and attached to every dot instead of placed at the boundary and relying on shuttling, which is challenging to achieve in a dense quantum dot array without mediators due to space constraints.

As with leakage reduction circuits, leakage \emph{detection} circuits~\cite{gottesmanStabilizerCodesQuantum,preskillFaulttolerantQuantumComputation1998} also require a signifincant increase in ancilla number or bring a significant cost in surface code runtime and circuit depth. A more practical approach would instead be to use physical charge detectors for leakage detection. In architectures without mediators, the leading leakage errors are one missing or one extra charge in the quantum dot. Charge detectors would therefore need to be interpersed within a densely packed quantum dot array and capable of accurately distinguish between the three different charge states. Furthermore, after the detection of a charge leakage, we cannot correct them by simply shuttling the leaked charge back because charge leakage can propagate across the array of quantum dots. Overall, this leads to significant practical challenges and spatial constraints. Furthermore,  the general leakage reduction/detection circuits described above assume the two-qubit gates in the leakage reduction/detection stage do not induce further leakage or transfer leakage and this is not the case for general two-qubit gates implemented in coupled quantum dot spins.

The practical challenges associated with integrating additional components or ancillae for leakage correction may be solved by using a modular structure~\cite{buonacorsiNetworkArchitectureTopological2019}. However, such a scheme creates a new source of leakage errors since it involves shuttling electrons across dozens of quantum dots. To keep the leakage error rate of \emph{across-array} shuttling low, we need to have an extremely low rate of \emph{between-dot} shuttling leakage, which means that we need to tune the gate voltages very slowly to maintain excellent adiabaticity. This leads to a trade-off between leakage suppression and the processing speed of the architecture. In addition, additional schemes to cope of leakage errors from the shuttling itself would be needed.

In architectures without mediators, if the parameters of direct exchange are chosen such that they have similar speed as mediated exchange, then the probability of leakage under direct exchange will be smaller than that using mediated exchange due to the higher energy of the excited charge state. However, if we did not take any active measures against the leakage errors, regardless of how small the leakage error probability is (as long as it is non-negligible), the leakages will keep accumulating until they break our code. As seen from above, active leakage correction schemes lead to a large runtime/qubit overhead. For a dense array of quantum dots, the reservoirs needed for leakage reset or the charge detectors needed for leakage detection are challenging to integrated due to space constraints, while in the modular scheme, the required electron shuttling creates a new source of leakage. In contrast, to handle leakage in our architecture, there are no additional components nor complex schemes required. We merely reset the mediators when they are idle, making our architecture more robust against charge leakage errors compared to the other quantum dot architectures.

\section{Resultant Computational Error from Leakage and Restoration}\label{app:leak_err_model}
For our system, there is no reason to assume that either the leakage event or the restoring charge relaxation are spin-conserving. Hence when a spin in a qubit dot is leaked and restored, we can assume that all the spin information is lost, which is equivalent to a depolarising error. When we look at the exchange interaction between qubit A and B via a mediator. If qubit A has leaked and been restored, it will be depolarised. Before the leakage, qubit B interacts with the original qubit A, and after the leakage qubit B interacts with the depolarised qubit A (via a mediator that might be faulty). The leakage and restoration can happen at any point during the exchange interaction, such uncertainty leads to a random depolarising error on the qubit B as well. 

Besides the depolarisation of the data qubit and the ancilla qubit involved in the exchange interaction, a leakage error may also lead to faulty mediator dots and hence affect the subsequent gates. Each stabiliser check cycle can be divided into two halves (interacting only inasmuch as they each include one dot of an ancilla double-dot pair): a five-dot system with one ancilla dot (A) connecting to two data dots (D1 and D2) via two mediators (M1 and M2). A interacts with D1 first via M1 in stage 1, then with D2 via M2 in stage 2. An error in  stage 1 only affects stage 2 if A has leaked and been restored using an electron from M2. In such a case, the left-over electron in M2 will be in a random state, thus when the electrons in A and D2 interact with the left over electrons in M2 in stage 2, they will also be depolarised regardless of whether further leakages and restorations happens in stage 2 or not.

Hence, we have the following leakage error table for the five-dot system with a exchange gate leakage probability $p$:
\begin{align*}
& \text{\qquad\ \  Stage 1  \ \ \qquad\qquad Stage 2 \qquad Errors}\\
&\begin{cases}
\parbox{6em}{\centering$1-p:$\\No leakages}&\begin{cases}
\parbox{6em}{\centering$1-p:$\\No leakages} &\text{Perfect}\\
\\
\parbox{6em}{\centering$p:$\\Any leakages} &\parbox{7em}{A and D2 depolarised}
\end{cases}\\
\\
\parbox{6em}{\centering$\frac{p}{4}: $\\\raggedright Leaked A and restored from M2}& \text{\qquad \quad Any \qquad \ \  All depolarised}\\
\\
\parbox{6.5em}{\centering$\frac{3p}{4}: $\\Other leakages}& \begin{cases}
\parbox{6em}{\centering$1-p:$\\No leakages} &\parbox{7em}{A and D1 depolarised}\\
\\
\parbox{6em}{\centering$p:$\\Any leakages} &\text{All depolarised}
\end{cases}\\
\end{cases}
\end{align*}
Hence, we can see here we have $(1-p)^2 = 1- 2p +p^2$ probability of having no leakage, and otherwise we will have partial or full depolarisation errors to the three qubits. In the calculations described in the main text, we have assumed an error model where we have $1-2p$ probability of having no leakage and otherwise have full depolarisation errors on all three qubits, which is a more severe error model than the more detailed one we describe here.

\section{One possible leakage mechanism}\label{app:leakage_mechanism}
\subsection{Three-dot system}
With reference to Figure~\ref{Fig:three_dot} in the main text, the charge configuration of the ground state is $(1, 2, 1)$. In the exchange Hamiltonian, the charge ground state is connected to the excited state charge states $(0, 3, 1)$ and $(1,3,0)$ via the tunnelling energies $t_{L2}$ and $t_{R2}$. Hence, the eigenstates of the exchange Hamiltonian are a superposition of the ground and excited state charge configurations. As noise causes such a superposition to decohere, there is a possibility that the exchange eigenstates will collapse into the excited state charge configuration, bringing the three dot setup out of the computational subspace, and leading to leakage errors. 

The probability of such a leakage error is related to the amplitude of the excited state in the coupled system eigenstate. Using perturbation theory, such an amplitude has a magnitude of $\frac{t}{\Delta}$, where $t$ is the tunnelling energy between the high energy state and the ground state while $\Delta$ is the energy difference between them. Hence, the possibility of our ground charge configuration $(1,2,1)$ escaping into the high energy charge configuration $(0, 3, 1)$, $(1,3,0)$ will be on the order of $\left(\frac{t_{L2}}{\Delta_L}\right)^2$ and $\left(\frac{t_{R2}}{\Delta_R}\right)^2$, which means that such leakage process will only be significant during the exchange interaction (when $\abs{\Delta_{L,R}}$ is small). Below we present a detailed analysis for the two-dot case, which can be easily generalised to our case. 
\subsection{Two-dot system}
\subsubsection{Hamiltonian}
For our two-dot system, we denote $\ket{T}$ as the triplet state with zero $z$-component, $\ket{S}$ as the singlet state, $\ket{\rm ion_+}$ as the state that has two electrons in one dot that can be reached by $\ket{S}$ via hopping, and $\ket{\rm ion_-}$ as the other state with two electrons in one dot but is orthogonal to $\ket{\rm ion_+}$.

We divide the Hamiltonian $H$ into two parts, a dominating diagonal part $H^{(0)}$:
\begin{align*}
H^{(0)} = \begin{pmatrix}
0&0&0&0\\0&0&0&0\\0&0&U&0\\0&0&0&U
\end{pmatrix}
\begin{matrix}
\ket{T} = \ket{0^{(0)}}\\\ket{S} = \ket{1^{(0)}}\\\ket{{\rm ion}_+} = \ket{2^{(0)}}\\\ket{{\rm ion}_-} = \ket{3^{(0)}}
\end{matrix}
\end{align*}
and a small off diagonal (tunnelling) part $r H^{(1)}$. 
\begin{align*}
rH^{(1)} = \begin{pmatrix}
0&0&0&0\\0&0&t+t^*&0\\0&t+t^*&0&0\\0&0&0&0
\end{pmatrix}
\begin{matrix}
\ket{T} = \ket{0^{(0)}}\\\ket{S} = \ket{1^{(0)}}\\\ket{{\rm ion}_+} = \ket{2^{(0)}}\\\ket{{\rm ion}_-} = \ket{3^{(0)}}
\end{matrix}
\end{align*}
$r$ here is the ratio between the off-diagonal tunnelling energy $t$ and the diagonal detuning energy $\Delta$:
\begin{align*}
r = \frac{t}{\Delta} \ll 1.
\end{align*}
Here we see that $rH^{(1)}$ only mixes $\ket{S}$ and $\ket{{\rm ion}_+}$ and leaves $\ket{T}$ and $\ket{{\rm ion}_-}$ unchanged. 

Starting from the eigenstates and the eigenenergies of $H^{(0)}$, we can obtain the eigenstates and the eigenenergies of $H$ using perturbation theory:
\begin{align*}
H  &= H^{(0)} + r H^{(1)}\\
\ket{n} &= \ket{n^{(0)}} + r\ket{n^{(1)}} + r^2\ket{n^{(2)}} + \cdots\\
E_{n} &= E_{n}^{(0)} + r E_{n}^{(1)} + r^2E_{n}^{(2)} + \cdots
\end{align*}
the superscript ${(m)}$ denotes the $m^{\text{th}}$-order correction.

\subsubsection{Perturbation theory}
\begin{itemize}
    \item Change in states $\Rightarrow$  \textbf{leakage error}:
    \begin{align}
    r\ket{1^{(1)}} & = \sum_{E_{n}^{(0)} \neq E_{1}^{(0)}} \ket{n^{(0)}} \frac{\bra{n^{(0)}} rH^{(1)} \ket{1^{(0)}}}{E_{1}^{(0)} - E_{n}^{(0)}} \nonumber\\
    & =  \ket{2^{(0)}}\frac{\bra{2^{(0)}} rH^{(1)}\ket{1^{(0)}}}{E_{1}^{(0)} - E_{2}^{(0)}}\nonumber\\
    & = -\frac{t+t^*}{U} \ket{2^{(0)}}\\
    r\ket{2^{(1)}} & = \sum_{E_{n}^{(0)} \neq E_{2}^{(0)}} \ket{n^{(0)}} \frac{\bra{n^{(0)}} rH^{(1)} \ket{2^{(0)}}}{E_{2}^{(0)} - E_{n}^{(0)}} \nonumber\\
    & =  \ket{1^{(0)}}\frac{\bra{1^{(0)}} rH^{(1)}\ket{2^{(0)}}}{E_{2}^{(0)} - E_{1}^{(0)}}\nonumber\\
    & = \frac{t+t^*}{U} \ket{1^{(0)}}
    \end{align}
    Hence
    \begin{align*}
    \ket{1} &= \ket{1^{(0)}} - \frac{t+t^*}{U} \ket{2^{(0)}}\\
    \ket{2} &= \ket{2^{(0)}} + \frac{t+t^*}{U} \ket{1^{(0)}}
    \end{align*}
    \item  Change in the ground state energy $\Rightarrow$ \textbf{exchange interaction}:
    
    The leading non-vanishing order of energy shift is
    \begin{align*}
    r^2E_1^{(2)} &= -2\frac{(t+t^*)^2}{U}\\
    r^2 E_2^{(2)}&= 2\frac{(t+t^*)^2}{U}
    \end{align*}
\end{itemize}

\subsubsection{Leakage oscillation}
Now if we start in the state of $\ket{S} = \ket{1^{(0)}}$ the probability of leaking into $\ket{\rm ion_+} = \ket{2^{(0)}}$ is:
\begin{align*}
&\quad \bra{2^{(0)}} e^{-i\hat{H}t}\ket{1^{(0)}} \\
&= \sum_ne^{-iE_nt} \braket{2^{(0)}}{n}\braket{n}{1^{(0)}}\\
& = e^{-iE_1t} \underbrace{\braket{2^{(0)}}{1}}_{-\frac{t+t^*}{U}} \underbrace{\braket{1}{1^{(0)}}}_{1} + e^{-iE_2t} \underbrace{\braket{2^{(0)}}{2}}_{1}\underbrace{\braket{2}{1^{(0)}}}_{\frac{t+t^*}{U}}\\
& = \frac{t+t^*}{U} \left(e^{-iE_2t} - e^{-iE_1t} \right)\\
& = \frac{t+t^*}{U} e^{-i\frac{E_2+E_1}{2}t} \left(e^{-i\frac{E_2 - E_1}{2}t} - e^{i\frac{E_2 - E_1}{2}t} \right)\\
& = \frac{t+t^*}{U} e^{-i\frac{E_2+E_1}{2}t} (-2i) \sin(\frac{E_2 - E_1}{2} t)
\end{align*}
Hence,
\begin{align*}
\abs{\bra{2^{(0)}} e^{-i\hat{H}t}\ket{1^{(0)}}}^2 &= 4\left(\frac{t+t^*}{U}\right)^2 \sin^2\left(\frac{E_2 - E_1}{2} t\right)
\end{align*}
To the leading order $E_2 - E_1 = U$. Hence, the probability of leaking has the magnitude of $r^2$ and oscillates with the frequency $\frac{U}{2}$
\begin{figure}[h]
    \centering
    \includegraphics[scale = 0.7]{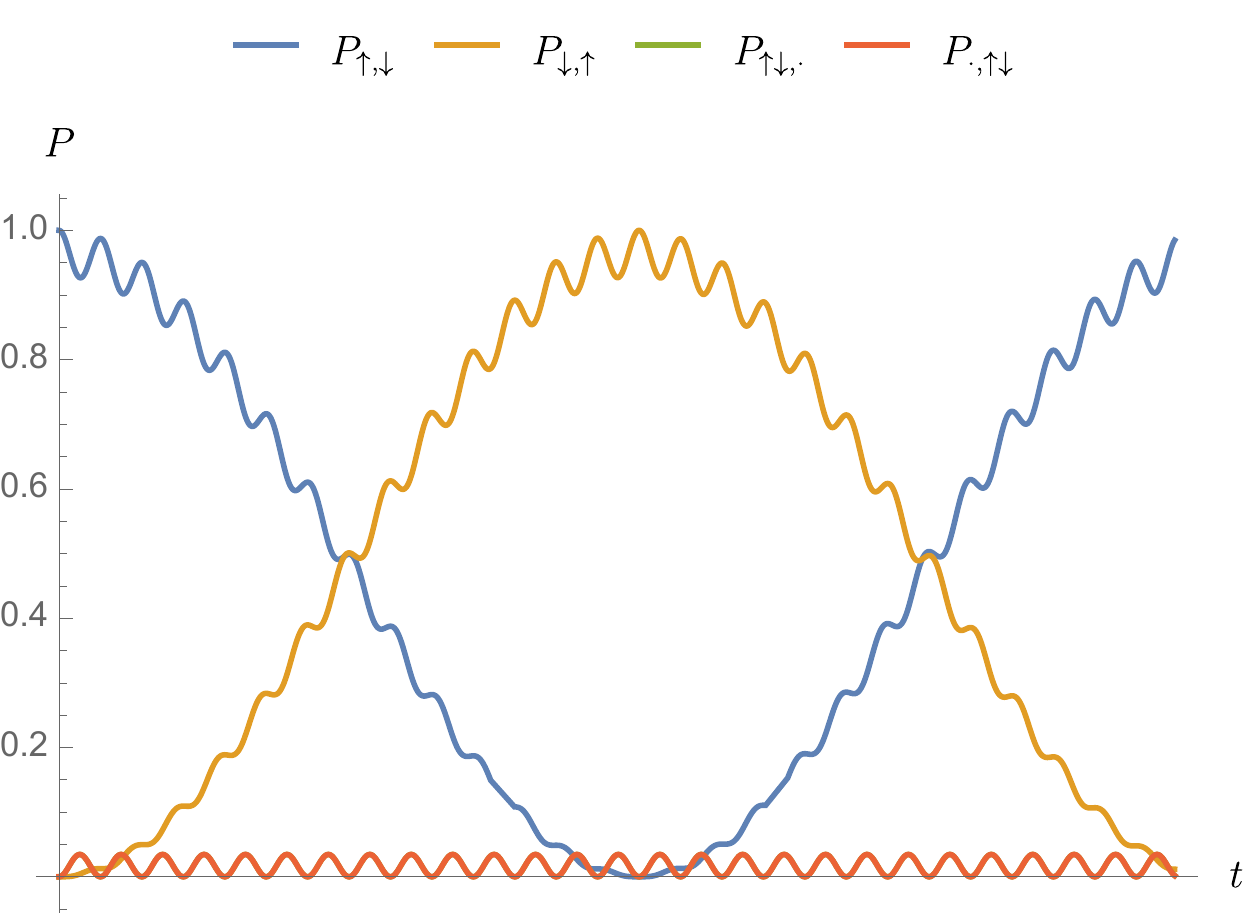}
    \caption{The probability of being in a different spin/charge states during one period of exchange interaction, following an initial $\ket{\uparrow, \downarrow}$ state. Note that the green and red lines completely overlap, and both represent a leakage probability. Here we have used $r = \frac{t}{\Delta} = 0.1$. }
    \label{Fig:two_dot}
\end{figure}
\begin{figure*}[ht]
    \centering
    \subfloat[With $S$ gate, $p_2 = 0.4\%$]{\includegraphics[width = 0.3\textwidth]{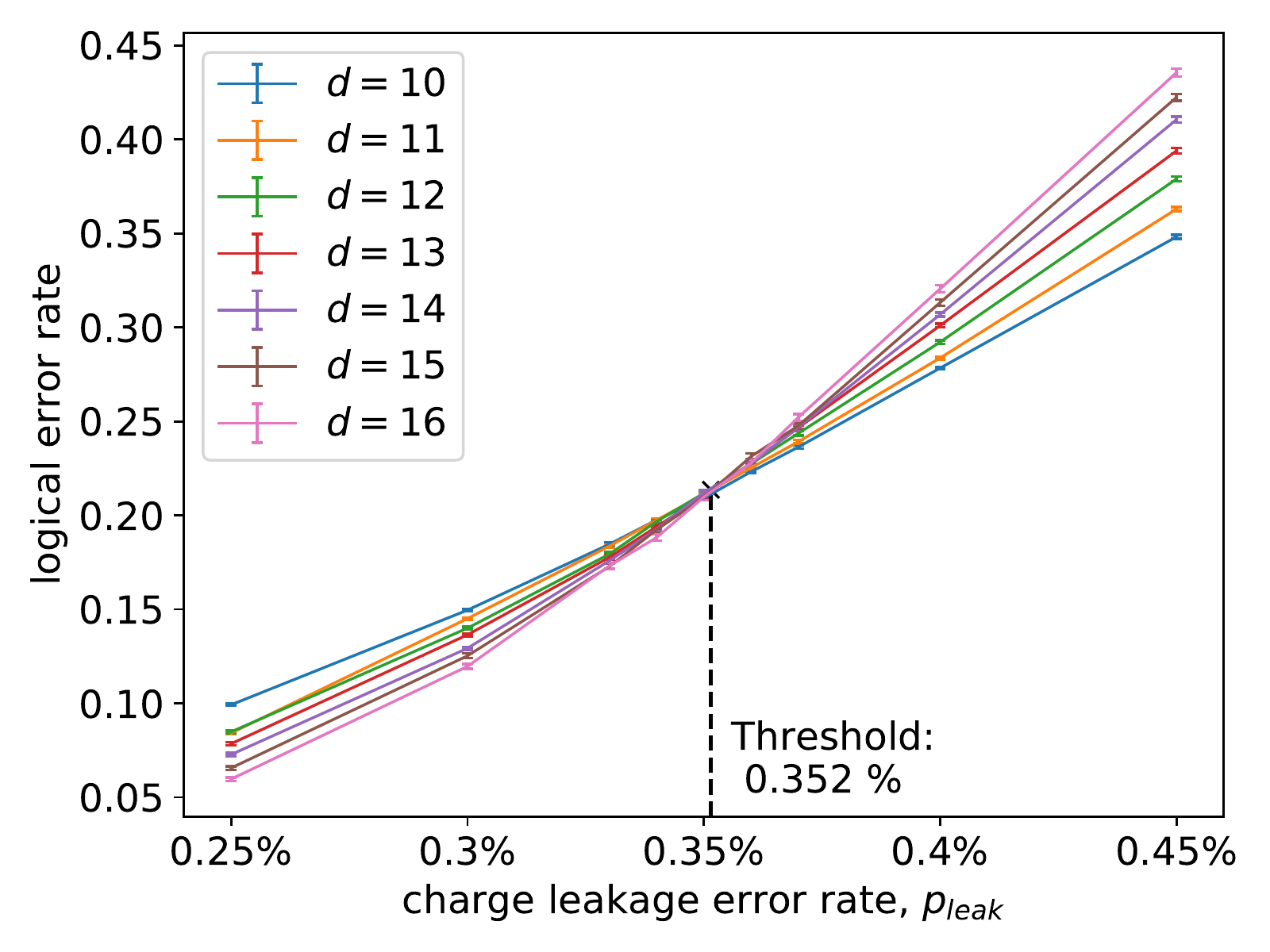}}
    \subfloat[With $\sqrt{\text{SWAP}}$ gate, $p_2 = 0.4\%$]{\includegraphics[width = 0.3\textwidth]{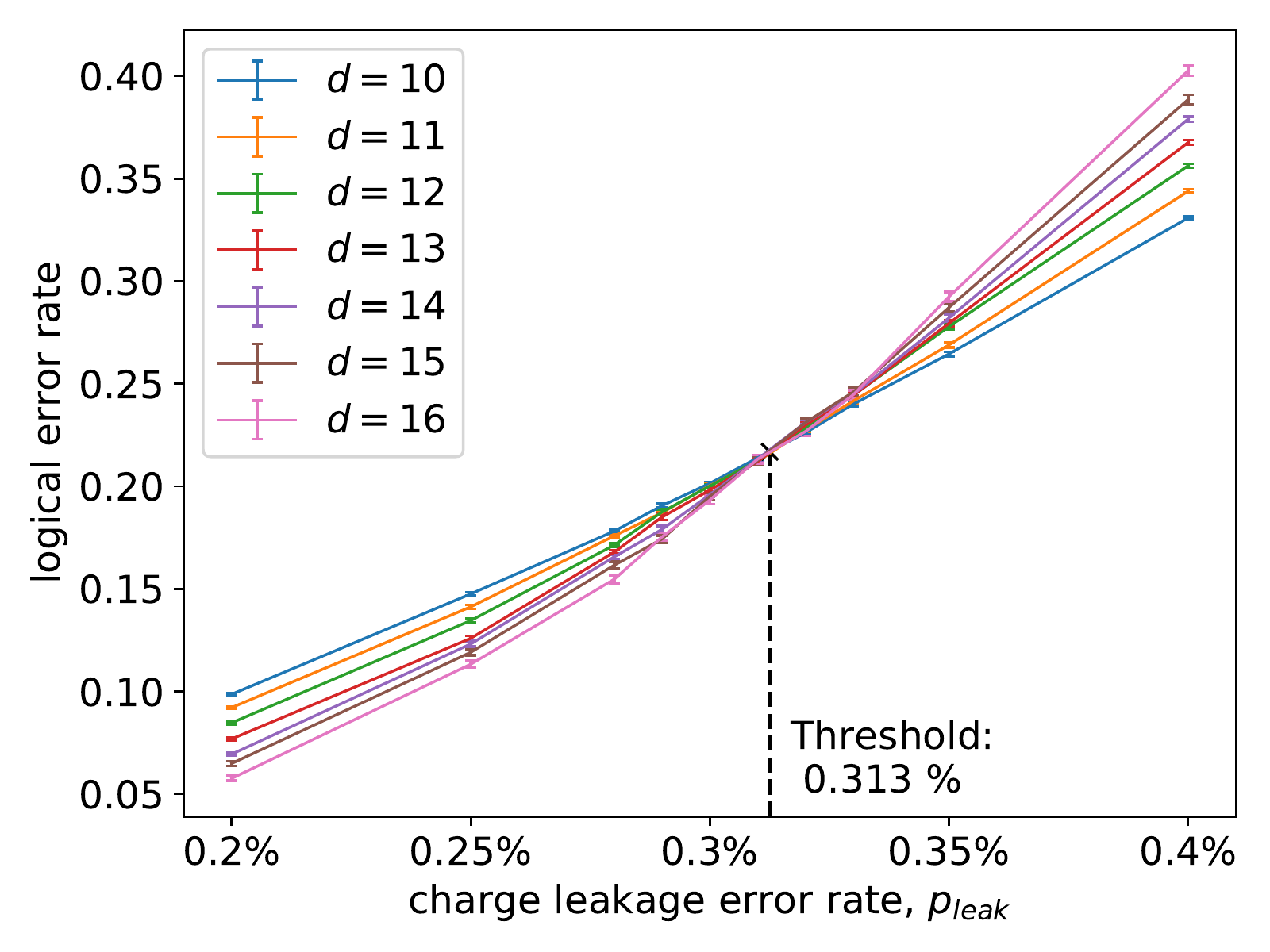}}
    \subfloat[With $S$ gate, $p_2 = 0.6\%$]{\includegraphics[width = 0.3\textwidth]{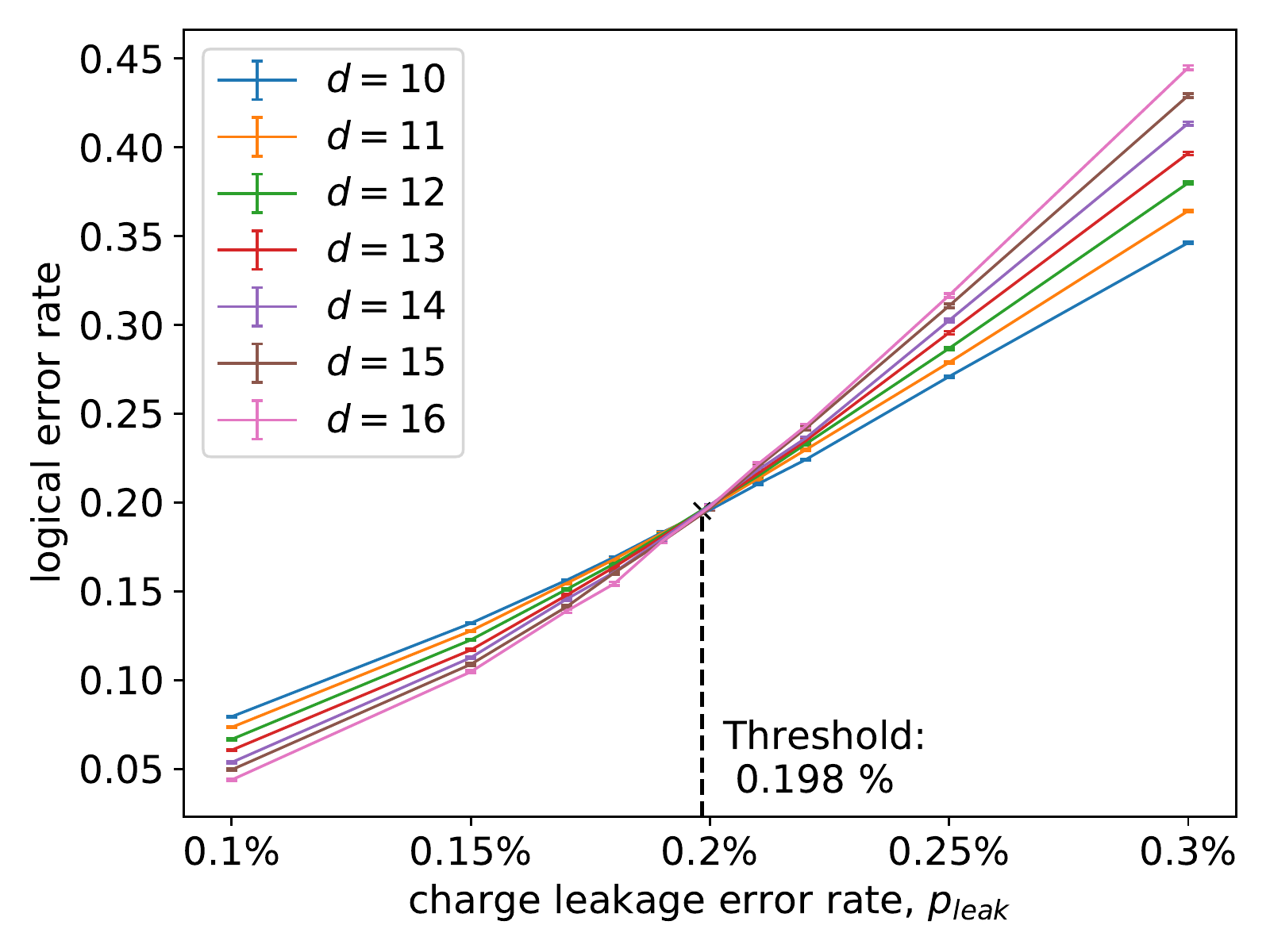}}\\
    \subfloat[With $\sqrt{\text{SWAP}}$ gate, $p_2 = 0.6\%$]{\includegraphics[width = 0.3\textwidth]{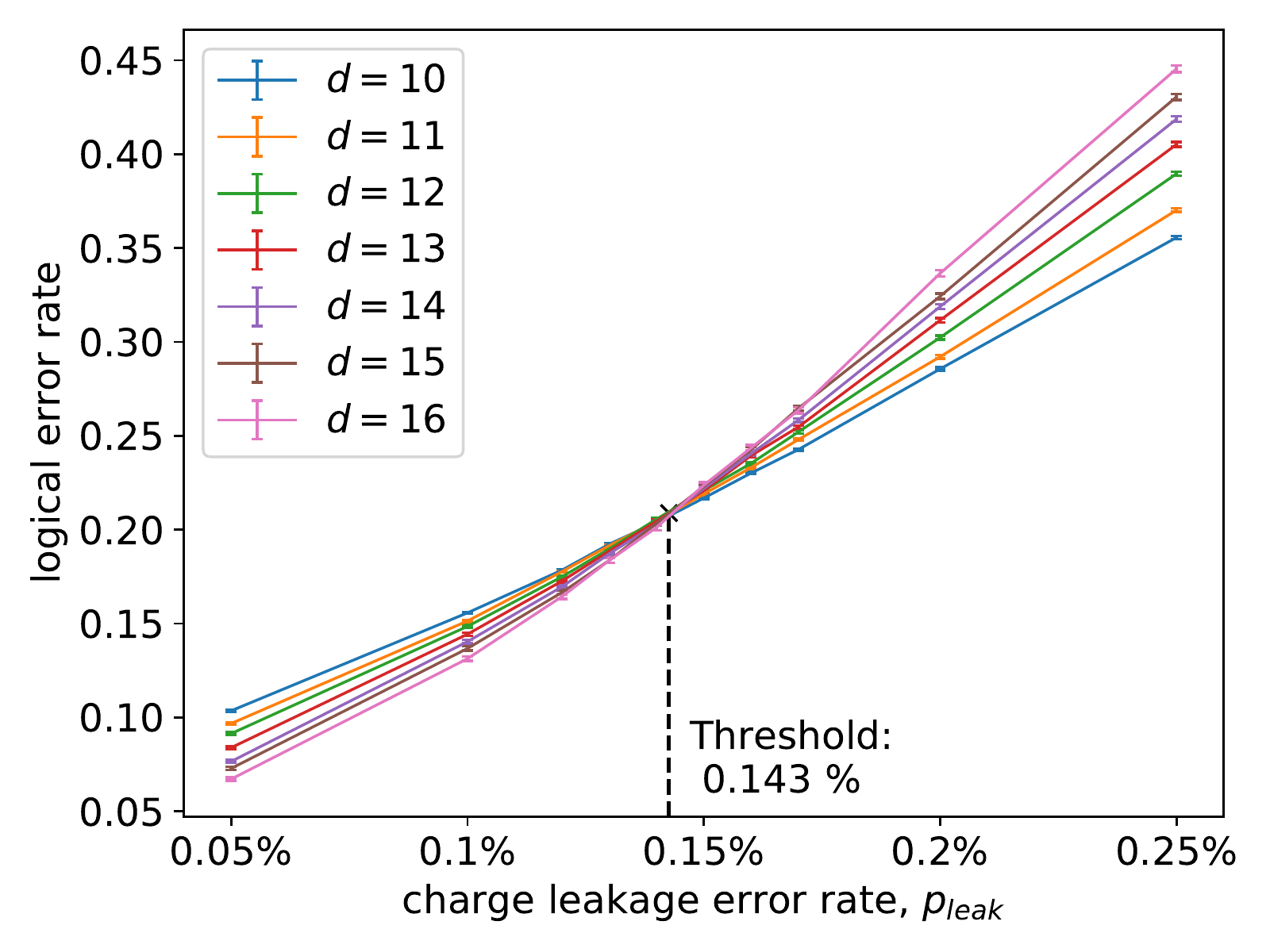}}
    \subfloat[With $S$ gate, $p_2 = 0.7\%$]{\includegraphics[width = 0.3\textwidth]{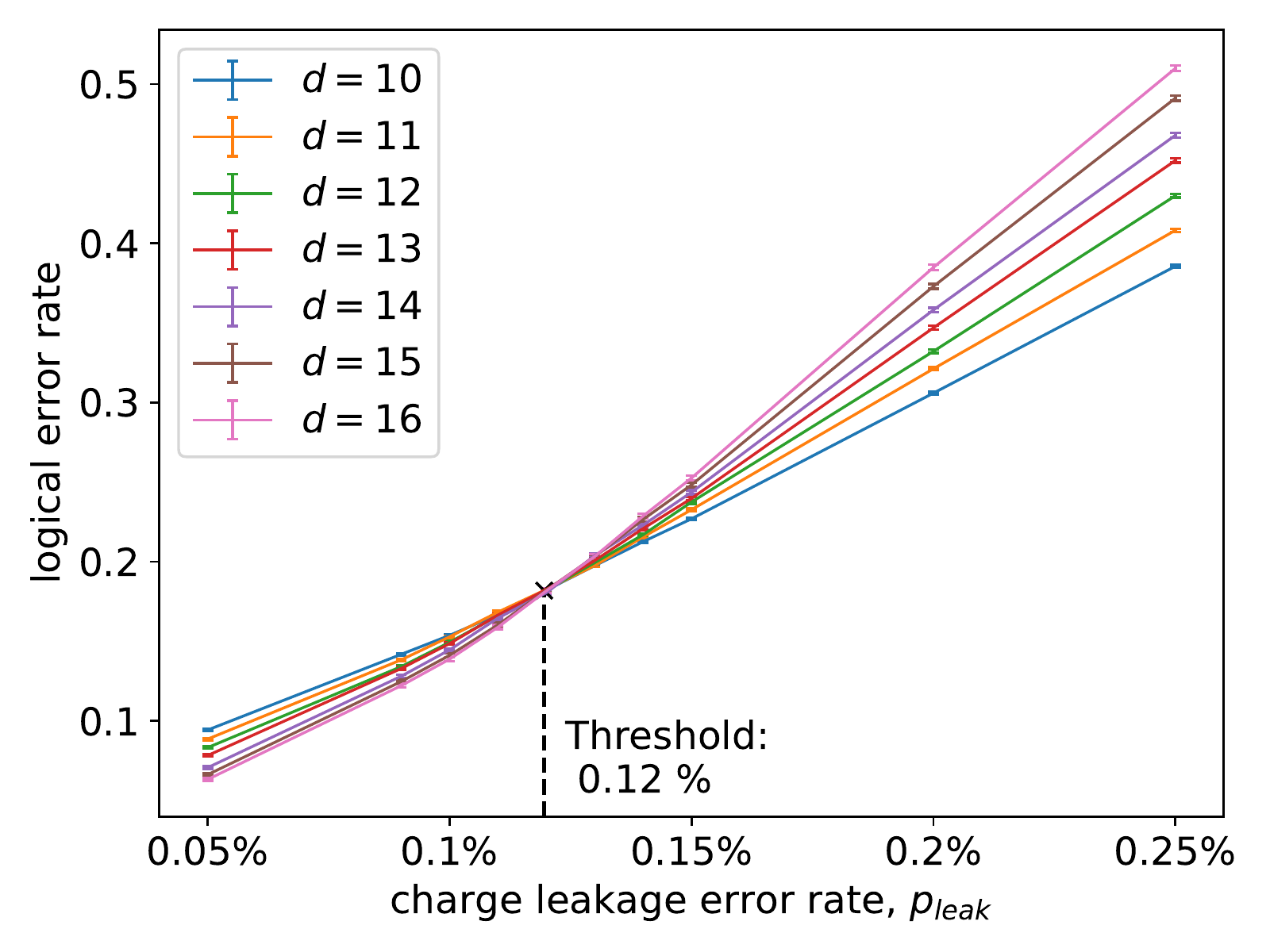}}
    \subfloat[With $\sqrt{\text{SWAP}}$ gate, $p_2 = 0.7\%$]{\includegraphics[width = 0.3\textwidth]{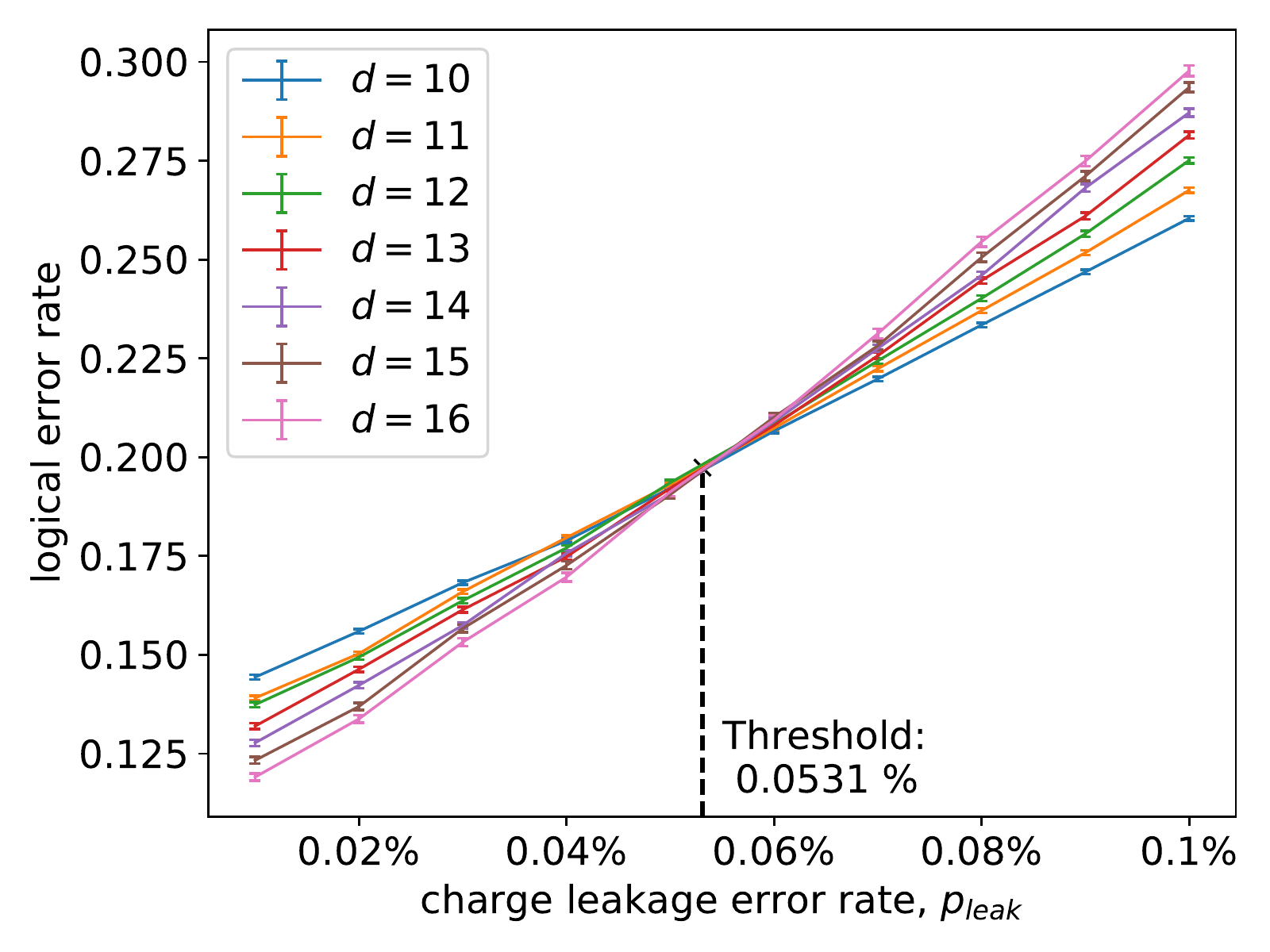}}
    \caption{Surface code leakage error $p_{\rm leak}$ threshold calculations assuming the use of $S$-gates with error probability $p_{2}$ or $\sqrt{\text{SWAP}}$ gates with error rate $p_2/2$. In all calculations, the error rate of single-qubit gates ($p_1$) and two-qubit gates ($p_2$) is assumed to be fixed $\left(\frac{p_1}{p_2} = 0.1\right)$. $d$ is the code distance of the surface code.}
    \label{Fig:more_threshold_plot}
\end{figure*}
\section{Threshold Simulation Details}
Based on the circuit and the error model outlined in Section~\ref{sect:surface_code}, we can obtain two error tables that outlines the probabilities of all possible error patterns (including both the errors on data qubits and the parity errors of the measurement results) when performing the X and Z stabiliser checks respectively. This will enable us to perform a Monte Carlo simulation of the stabiliser check process with errors arising according to the probability obtained from the error tables. Each round of stabiliser checks will give rise to a 2D grid of parity check results. For a distance-$d$ surface code, we will repeat our stabiliser measurement for $d$ times to fight with measurement errors, which can be viewed as stacking up $d$ layers of 2D parity result grid, giving rise to a 3D grid with one of the dimension being time~\cite{stephensFaulttolerantThresholdsQuantum2014}. We can then try to match the failed parity checks to the boundary or to any change in the parity results in the time direction using minimum-weight perfect matching(MWPM), which is carried out using the Blossom V package~\cite{kolmogorovBlossomNewImplementation2009}. The surface code threshold simulation module we used is published on Github~\cite{HttpsGithubCom}. For simplicity, in our simulation we have the same weight for the edges in the spatial direction and the edges in the time direction. However, threshold improvements can be gained by optimising the weight ratios between them due to the different probabilities of failure. Further improvements of the threshold can be achieved by using more advance decoders, e.g. the maximum likelihood decoder~\cite{bravyiEfficientAlgorithmsMaximum2014}.

\section{Charge Leakage Threshold Under Different Gate Error Rates}\label{sect:further_sim}
Table~\ref{table:02}  here summarise the charge leakage ($p_{\rm leak}$) threshold under a range of different gate error ($p_2$), with the additional threshold plots shown in Figure~\ref{Fig:more_threshold_plot}.

\begin{table}[htbp]
    \centering
    \begin{tabular}{l|  c r}
        $p_2$ & $S$-gate & $\sqrt{\text{SWAP}}$\\
        \hline
        $0 \%$ &\multicolumn{2}{c}{0.66}\\
        $0.4\%$ & $0.35\%$      & $0.31\%$\\
        $0.5\%$ & $0.27\%$      & $0.23\%$\\
        $0.6\%$ & $0.20\%$      & $0.14\%$\\
        $0.7\%$ & $0.12\%$		& $0.05\%$\\
        $0.76\%$ & -- & $0\%$\\
        $0.86\%$ & $0\%$ & nil\\
    \end{tabular}
    \caption{The charge leakage ($p_{\rm leak}$) threshold under different gate error rate ($p_2$), assuming the use of $S$-gates with error rate $p_{2}$ or $\sqrt{\text{SWAP}}$ gates with error rate $p_2/2$. }
    \label{table:02}
\end{table}
%\newpage
% Note that placing {figure*} right before the biblio may lead to errors, thus we have move the figure here somewhere else.

%\bibliographystyle{unsrtnat}
%\bibliography{ref}

\begin{thebibliography}{90}
\providecommand{\natexlab}[1]{#1}
\providecommand{\url}[1]{\texttt{#1}}
\expandafter\ifx\csname urlstyle\endcsname\relax
  \providecommand{\doi}[1]{doi: #1}\else
  \providecommand{\doi}{doi: \begingroup \urlstyle{rm}\Url}\fi

\bibitem[Feynman(1982)]{feynmanSimulatingPhysicsComputers1982}
Richard~P. Feynman.
\newblock Simulating physics with computers.
\newblock \emph{International Journal of Theoretical Physics}, 21\penalty0
  (6-7):\penalty0 467--488, June 1982.
\newblock \doi{10.1007/BF02650179}.

\bibitem[Grover(1996)]{groverFastQuantumMechanical1996}
Lov~K. Grover.
\newblock A {{Fast Quantum Mechanical Algorithm}} for {{Database Search}}.
\newblock In \emph{Proceedings of the {{Twenty}}-Eighth {{Annual ACM
  Symposium}} on {{Theory}} of {{Computing}}}, {{STOC}} '96, pages 212--219,
  {New York, NY, USA}, 1996. {ACM}.
\newblock \doi{10.1145/237814.237866}.

\bibitem[Grover(1997)]{groverQuantumMechanicsHelps1997}
Lov~K. Grover.
\newblock Quantum {{Mechanics Helps}} in {{Searching}} for a {{Needle}} in a
  {{Haystack}}.
\newblock \emph{Physical Review Letters}, 79\penalty0 (2):\penalty0 325--328,
  July 1997.
\newblock \doi{10.1103/PhysRevLett.79.325}.

\bibitem[Temme et~al.(2011)Temme, Osborne, Vollbrecht, Poulin, and
  Verstraete]{temmeQuantumMetropolisSampling2011}
K.~Temme, T.~J. Osborne, K.~G. Vollbrecht, D.~Poulin, and F.~Verstraete.
\newblock Quantum {{Metropolis}} sampling.
\newblock \emph{Nature}, 471\penalty0 (7336):\penalty0 87--90, March 2011.
\newblock \doi{10.1038/nature09770}.

\bibitem[{Montanaro Ashley}(2015)]{montanaroashleyQuantumSpeedupMonte2015}
{Montanaro Ashley}.
\newblock Quantum speedup of {{Monte Carlo}} methods.
\newblock \emph{Proceedings of the Royal Society A: Mathematical, Physical and
  Engineering Sciences}, 471\penalty0 (2181):\penalty0 20150301, September
  2015.
\newblock \doi{10.1098/rspa.2015.0301}.

\bibitem[Wang et~al.(2011)Wang, Fowler, and
  Hollenberg]{wangSurfaceCodeQuantum2011}
David~S. Wang, Austin~G. Fowler, and Lloyd C.~L. Hollenberg.
\newblock Surface code quantum computing with error rates over 1\%.
\newblock \emph{Physical Review A}, 83\penalty0 (2), February 2011.
\newblock \doi{10.1103/PhysRevA.83.020302}.

\bibitem[Fowler et~al.(2012)Fowler, Mariantoni, Martinis, and
  Cleland]{fowlerSurfaceCodesPractical2012}
Austin~G. Fowler, Matteo Mariantoni, John~M. Martinis, and Andrew~N. Cleland.
\newblock Surface codes: {{Towards}} practical large-scale quantum computation.
\newblock \emph{Physical Review A}, 86\penalty0 (3):\penalty0 032324, 2012.
\newblock \doi{10.1103/PhysRevA.86.032324}.

\bibitem[Lekitsch et~al.(2017)Lekitsch, Weidt, Fowler, M{\o}lmer, Devitt,
  Wunderlich, and Hensinger]{lekitschBlueprintMicrowaveTrapped2017}
Bjoern Lekitsch, Sebastian Weidt, Austin~G. Fowler, Klaus M{\o}lmer, Simon~J.
  Devitt, Christof Wunderlich, and Winfried~K. Hensinger.
\newblock Blueprint for a microwave trapped ion quantum computer.
\newblock \emph{Science Advances}, 3\penalty0 (2):\penalty0 e1601540, February
  2017.
\newblock \doi{10.1126/sciadv.1601540}.

\bibitem[Hill et~al.(2015)Hill, Peretz, Hile, House, Fuechsle, Rogge, Simmons,
  and Hollenberg]{hillSurfaceCodeQuantum2015}
C.~D. Hill, E.~Peretz, S.~J. Hile, M.~G. House, M.~Fuechsle, S.~Rogge, M.~Y.
  Simmons, and L.~C.~L. Hollenberg.
\newblock A surface code quantum computer in silicon.
\newblock \emph{Science Advances}, 1\penalty0 (9):\penalty0 e1500707--e1500707,
  October 2015.
\newblock \doi{10.1126/sciadv.1500707}.

\bibitem[O'Gorman et~al.(2016)O'Gorman, Nickerson, Ross, Morton, and
  Benjamin]{ogormanSiliconbasedSurfaceCode2016}
Joe O'Gorman, Naomi~H. Nickerson, Philipp Ross, John~JL Morton, and Simon~C.
  Benjamin.
\newblock A silicon-based surface code quantum computer.
\newblock \emph{npj Quantum Information}, 2:\penalty0 npjqi201519, February
  2016.
\newblock \doi{10.1038/npjqi.2015.19}.

\bibitem[O'Gorman and Campbell(2017)]{ogormanQuantumComputationRealistic2017}
Joe O'Gorman and Earl~T. Campbell.
\newblock Quantum computation with realistic magic-state factories.
\newblock \emph{Physical Review A}, 95\penalty0 (3), March 2017.
\newblock \doi{10.1103/PhysRevA.95.032338}.

\bibitem[Vandersypen et~al.(2017)Vandersypen, Bluhm, Clarke, Dzurak, Ishihara,
  Morello, Reilly, Schreiber, and
  Veldhorst]{vandersypenInterfacingSpinQubits2017}
L.~M.~K. Vandersypen, H.~Bluhm, J.~S. Clarke, A.~S. Dzurak, R.~Ishihara,
  A.~Morello, D.~J. Reilly, L.~R. Schreiber, and M.~Veldhorst.
\newblock Interfacing spin qubits in quantum dots and donors\textemdash{}hot,
  dense, and coherent.
\newblock \emph{npj Quantum Information}, 3\penalty0 (1), December 2017.
\newblock \doi{10.1038/s41534-017-0038-y}.

\bibitem[Veldhorst et~al.(2017)Veldhorst, Eenink, Yang, and
  Dzurak]{veldhorstSiliconCMOSArchitecture2017}
M.~Veldhorst, H.~G.~J. Eenink, C.~H. Yang, and A.~S. Dzurak.
\newblock Silicon {{CMOS}} architecture for a spin-based quantum computer.
\newblock \emph{Nature Communications}, 8\penalty0 (1):\penalty0 1766, December
  2017.
\newblock \doi{10.1038/s41467-017-01905-6}.

\bibitem[Li et~al.(2018)Li, Petit, Franke, Dehollain, Helsen, Steudtner,
  Thomas, Yoscovits, Singh, Wehner, Vandersypen, Clarke, and
  Veldhorst]{liCrossbarNetworkSilicon2018}
Ruoyu Li, Luca Petit, David~P. Franke, Juan~Pablo Dehollain, Jonas Helsen, Mark
  Steudtner, Nicole~K. Thomas, Zachary~R. Yoscovits, Kanwal~J. Singh, Stephanie
  Wehner, Lieven M.~K. Vandersypen, James~S. Clarke, and Menno Veldhorst.
\newblock A crossbar network for silicon quantum dot qubits.
\newblock \emph{Science Advances}, 4\penalty0 (7):\penalty0 eaar3960, July
  2018.
\newblock \doi{10.1126/sciadv.aar3960}.

\bibitem[Buonacorsi et~al.(2019)Buonacorsi, Cai, Ramirez, Willick, Walker, Li,
  Shaw, Xu, Benjamin, and Baugh]{buonacorsiNetworkArchitectureTopological2019}
Brandon Buonacorsi, Zhenyu Cai, Eduardo~B. Ramirez, Kyle~S. Willick, Sean~M.
  Walker, Jiahao Li, Benjamin~D. Shaw, Xiaosi Xu, Simon~C. Benjamin, and
  Jonathan Baugh.
\newblock Network architecture for a topological quantum computer in silicon.
\newblock \emph{Quantum Science and Technology}, 4\penalty0 (2):\penalty0
  025003, January 2019.
\newblock \doi{10.1088/2058-9565/aaf3c4}.

\bibitem[Motzoi et~al.(2009)Motzoi, Gambetta, Rebentrost, and
  Wilhelm]{motzoiSimplePulsesElimination2009}
F.~Motzoi, J.~M. Gambetta, P.~Rebentrost, and F.~K. Wilhelm.
\newblock Simple {{Pulses}} for {{Elimination}} of {{Leakage}} in {{Weakly
  Nonlinear Qubits}}.
\newblock \emph{Physical Review Letters}, 103\penalty0 (11):\penalty0 110501,
  September 2009.
\newblock \doi{10.1103/PhysRevLett.103.110501}.

\bibitem[Ferr{\'o}n and
  Dom{\'i}nguez(2010)]{ferronIntrinsicLeakageJosephson2010}
Alejandro Ferr{\'o}n and Daniel Dom{\'i}nguez.
\newblock Intrinsic leakage of the {{Josephson}} flux qubit and breakdown of
  the two-level approximation for strong driving.
\newblock \emph{Physical Review B}, 81\penalty0 (10):\penalty0 104505, March
  2010.
\newblock \doi{10.1103/PhysRevB.81.104505}.

\bibitem[Duan et~al.(2001)Duan, Cirac, and
  Zoller]{duanGeometricManipulationTrapped2001}
L.-M. Duan, J.~I. Cirac, and P.~Zoller.
\newblock Geometric {{Manipulation}} of {{Trapped Ions}} for {{Quantum
  Computation}}.
\newblock \emph{Science}, 292\penalty0 (5522):\penalty0 1695--1697, June 2001.
\newblock \doi{10.1126/science.1058835}.

\bibitem[Haffner et~al.(2008)Haffner, Roos, and
  Blatt]{haffnerQuantumComputingTrapped2008}
H~Haffner, C~Roos, and R~Blatt.
\newblock Quantum computing with trapped ions.
\newblock \emph{Physics Reports}, 469\penalty0 (4):\penalty0 155--203, December
  2008.
\newblock \doi{10.1016/j.physrep.2008.09.003}.

\bibitem[Fong and Wandzura(2011)]{fongUniversalQuantumComputation2011}
Bryan~H. Fong and Stephen~M. Wandzura.
\newblock Universal {{Quantum Computation}} and {{Leakage Reduction}} in the
  3-{{Qubit Decoherence Free Subsystem}}.
\newblock February 2011.
\newblock URL \url{https://arxiv.org/abs/1102.2909v1}.

\bibitem[Mehl et~al.(2015)Mehl, Bluhm, and
  DiVincenzo]{mehlFaulttolerantQuantumComputation2015}
Sebastian Mehl, Hendrik Bluhm, and David~P. DiVincenzo.
\newblock Fault-tolerant quantum computation for singlet-triplet qubits with
  leakage errors.
\newblock \emph{Physical Review B}, 91\penalty0 (8):\penalty0 085419, February
  2015.
\newblock \doi{10.1103/PhysRevB.91.085419}.

\bibitem[Malinowski et~al.(2019)Malinowski, Martins, Smith, Bartlett, Doherty,
  Nissen, Fallahi, Gardner, Manfra, Marcus, and
  Kuemmeth]{malinowskiFastSpinExchange2019}
Filip~K. Malinowski, Frederico Martins, Thomas~B. Smith, Stephen~D. Bartlett,
  Andrew~C. Doherty, Peter~D. Nissen, Saeed Fallahi, Geoffrey~C. Gardner,
  Michael~J. Manfra, Charles~M. Marcus, and Ferdinand Kuemmeth.
\newblock Fast spin exchange across a multielectron mediator.
\newblock \emph{Nature Communications}, 10\penalty0 (1):\penalty0 1196, March
  2019.
\newblock \doi{10.1038/s41467-019-09194-x}.

\bibitem[Angus et~al.(2007)Angus, Ferguson, Dzurak, and
  Clark]{angusGateDefinedQuantumDots2007}
Susan~J. Angus, Andrew~J. Ferguson, Andrew~S. Dzurak, and Robert~G. Clark.
\newblock Gate-{{Defined Quantum Dots}} in {{Intrinsic Silicon}}.
\newblock \emph{Nano Letters}, 7\penalty0 (7):\penalty0 2051--2055, July 2007.
\newblock \doi{10.1021/nl070949k}.

\bibitem[Veldhorst et~al.(2014)Veldhorst, Hwang, Yang, Leenstra, {de Ronde},
  Dehollain, Muhonen, Hudson, Itoh, Morello, and
  Dzurak]{veldhorstAddressableQuantumDot2014}
M.~Veldhorst, J.~C.~C. Hwang, C.~H. Yang, A.~W. Leenstra, B.~{de Ronde}, J.~P.
  Dehollain, J.~T. Muhonen, F.~E. Hudson, K.~M. Itoh, A.~Morello, and A.~S.
  Dzurak.
\newblock An addressable quantum dot qubit with fault-tolerant
  control-fidelity.
\newblock \emph{Nature Nanotechnology}, 9\penalty0 (12):\penalty0 981--985,
  December 2014.
\newblock \doi{10.1038/nnano.2014.216}.

\bibitem[Chan et~al.(2018)Chan, Huang, Yang, Hwang, Hensen, Tanttu, Hudson,
  Itoh, Laucht, Morello, and Dzurak]{chanAssessmentSiliconQuantum2018}
K.~W. Chan, W.~Huang, C.~H. Yang, J.~C.~C. Hwang, B.~Hensen, T.~Tanttu, F.~E.
  Hudson, K.~M. Itoh, A.~Laucht, A.~Morello, and A.~S. Dzurak.
\newblock Assessment of a silicon quantum dot spin qubit environment via noise
  spectroscopy.
\newblock \emph{Physical Review Applied}, 10\penalty0 (4), October 2018.
\newblock \doi{10.1103/PhysRevApplied.10.044017}.

\bibitem[Yang et~al.(2019)Yang, Chan, Harper, Huang, Evans, Hwang, Hensen,
  Laucht, Tanttu, Hudson, Flammia, Itoh, Morello, Bartlett, and
  Dzurak]{yangSiliconQubitFidelities2019}
C.~H. Yang, K.~W. Chan, R.~Harper, W.~Huang, T.~Evans, J.~C.~C. Hwang,
  B.~Hensen, A.~Laucht, T.~Tanttu, F.~E. Hudson, S.~T. Flammia, K.~M. Itoh,
  A.~Morello, S.~D. Bartlett, and A.~S. Dzurak.
\newblock Silicon qubit fidelities approaching incoherent noise limits via
  pulse engineering.
\newblock \emph{Nature Electronics}, 2\penalty0 (4):\penalty0 151, April 2019.
\newblock \doi{10.1038/s41928-019-0234-1}.

\bibitem[Kawakami et~al.(2016)Kawakami, Jullien, Scarlino, Ward, Savage,
  Lagally, Dobrovitski, Friesen, Coppersmith, Eriksson, and
  Vandersypen]{kawakamiGateFidelityCoherence2016}
Erika Kawakami, Thibaut Jullien, Pasquale Scarlino, Daniel~R. Ward, Donald~E.
  Savage, Max~G. Lagally, Viatcheslav~V. Dobrovitski, Mark Friesen, Susan~N.
  Coppersmith, Mark~A. Eriksson, and Lieven M.~K. Vandersypen.
\newblock Gate fidelity and coherence of an electron spin in an {{Si}}/{{SiGe}}
  quantum dot with micromagnet.
\newblock \emph{Proceedings of the National Academy of Sciences}, 113\penalty0
  (42):\penalty0 11738--11743, October 2016.
\newblock \doi{10.1073/pnas.1603251113}.

\bibitem[Yoneda et~al.(2018)Yoneda, Takeda, Otsuka, Nakajima, Delbecq, Allison,
  Honda, Kodera, Oda, Hoshi, Usami, Itoh, and
  Tarucha]{yonedaQuantumdotSpinQubit2018}
Jun Yoneda, Kenta Takeda, Tomohiro Otsuka, Takashi Nakajima, Matthieu~R.
  Delbecq, Giles Allison, Takumu Honda, Tetsuo Kodera, Shunri Oda, Yusuke
  Hoshi, Noritaka Usami, Kohei~M. Itoh, and Seigo Tarucha.
\newblock A quantum-dot spin qubit with coherence limited by charge noise and
  fidelity higher than 99.9\%.
\newblock \emph{Nature Nanotechnology}, 13\penalty0 (2):\penalty0 102--106,
  February 2018.
\newblock \doi{10.1038/s41565-017-0014-x}.

\bibitem[Pica et~al.(2016)Pica, Lovett, Bhatt, Schenkel, and
  Lyon]{picaSurfaceCodeArchitecture2016}
G.~Pica, B.~W. Lovett, R.~N. Bhatt, T.~Schenkel, and S.~A. Lyon.
\newblock Surface code architecture for donors and dots in silicon with
  imprecise and nonuniform qubit couplings.
\newblock \emph{Physical Review B}, 93\penalty0 (3), January 2016.
\newblock \doi{10.1103/PhysRevB.93.035306}.

\bibitem[Jones et~al.(2018)Jones, Fogarty, Morello, Gyure, Dzurak, and
  Ladd]{jonesLogicalQubitLinear2018}
Cody Jones, Michael~A. Fogarty, Andrea Morello, Mark~F. Gyure, Andrew~S.
  Dzurak, and Thaddeus~D. Ladd.
\newblock Logical {{Qubit}} in a {{Linear Array}} of {{Semiconductor Quantum
  Dots}}.
\newblock \emph{Physical Review X}, 8\penalty0 (2):\penalty0 021058, June 2018.
\newblock \doi{10.1103/PhysRevX.8.021058}.

\bibitem[Tyryshkin et~al.(2012)Tyryshkin, Tojo, Morton, Riemann, Abrosimov,
  Becker, Pohl, Schenkel, Thewalt, Itoh, and
  Lyon]{tyryshkinElectronSpinCoherence2012}
Alexei~M. Tyryshkin, Shinichi Tojo, John J.~L. Morton, Helge Riemann,
  Nikolai~V. Abrosimov, Peter Becker, Hans-Joachim Pohl, Thomas Schenkel,
  Michael L.~W. Thewalt, Kohei~M. Itoh, and S.~A. Lyon.
\newblock Electron spin coherence exceeding seconds in high-purity silicon.
\newblock \emph{Nature Materials}, 11\penalty0 (2):\penalty0 143--147, February
  2012.
\newblock \doi{10.1038/nmat3182}.

\bibitem[Laucht et~al.(2015)Laucht, Muhonen, Mohiyaddin, Kalra, Dehollain,
  Freer, Hudson, Veldhorst, Rahman, Klimeck, Itoh, Jamieson, McCallum, Dzurak,
  and Morello]{lauchtElectricallyControllingSinglespin2015}
Arne Laucht, Juha~T. Muhonen, Fahd~A. Mohiyaddin, Rachpon Kalra, Juan~P.
  Dehollain, Solomon Freer, Fay~E. Hudson, Menno Veldhorst, Rajib Rahman,
  Gerhard Klimeck, Kohei~M. Itoh, David~N. Jamieson, Jeffrey~C. McCallum,
  Andrew~S. Dzurak, and Andrea Morello.
\newblock Electrically controlling single-spin qubits in a continuous microwave
  field.
\newblock \emph{Science Advances}, 1\penalty0 (3):\penalty0 e1500022, April
  2015.
\newblock \doi{10.1126/sciadv.1500022}.

\bibitem[Rist{\`e} et~al.(2015)Rist{\`e}, Poletto, Huang, Bruno, Vesterinen,
  Saira, and DiCarlo]{risteDetectingBitflipErrors2015}
D.~Rist{\`e}, S.~Poletto, M.-Z. Huang, A.~Bruno, V.~Vesterinen, O.-P. Saira,
  and L.~DiCarlo.
\newblock Detecting bit-flip errors in a logical qubit using stabilizer
  measurements.
\newblock \emph{Nature Communications}, 6:\penalty0 6983, April 2015.
\newblock \doi{10.1038/ncomms7983}.

\bibitem[Schutjens et~al.(2013)Schutjens, Dagga, Egger, and
  Wilhelm]{schutjensSinglequbitGatesFrequencycrowded2013}
R.~Schutjens, F.~Abu Dagga, D.~J. Egger, and F.~K. Wilhelm.
\newblock Single-qubit gates in frequency-crowded transmon systems.
\newblock \emph{Physical Review A}, 88\penalty0 (5):\penalty0 052330, November
  2013.
\newblock \doi{10.1103/PhysRevA.88.052330}.

\bibitem[Yang et~al.(2013)Yang, Rossi, Ruskov, Lai, Mohiyaddin, Lee, Tahan,
  Klimeck, Morello, and Dzurak]{yangSpinvalleyLifetimesSilicon2013}
C.~H. Yang, A.~Rossi, R.~Ruskov, N.~S. Lai, F.~A. Mohiyaddin, S.~Lee, C.~Tahan,
  G.~Klimeck, A.~Morello, and A.~S. Dzurak.
\newblock Spin-valley lifetimes in a silicon quantum dot with tunable valley
  splitting.
\newblock \emph{Nature Communications}, 4:\penalty0 2069, June 2013.
\newblock \doi{10.1038/ncomms3069}.

\bibitem[Kawakami et~al.(2014)Kawakami, Scarlino, Ward, Braakman, Savage,
  Lagally, Friesen, Coppersmith, Eriksson, and
  Vandersypen]{kawakamiElectricalControlLonglived2014}
E.~Kawakami, P.~Scarlino, D.~R. Ward, F.~R. Braakman, D.~E. Savage, M.~G.
  Lagally, Mark Friesen, S.~N. Coppersmith, M.~A. Eriksson, and L.~M.~K.
  Vandersypen.
\newblock Electrical control of a long-lived spin qubit in a {{Si}}/{{SiGe}}
  quantum dot.
\newblock \emph{Nature Nanotechnology}, 9\penalty0 (9):\penalty0 666--670,
  September 2014.
\newblock \doi{10.1038/nnano.2014.153}.

\bibitem[Leon et~al.(2019)Leon, Yang, Hwang, Lemyre, Tanttu, Huang, Chan, Tan,
  Hudson, Itoh, Morello, Laucht, {Pioro-Ladriere}, Saraiva, and
  Dzurak]{leonCoherentSpinControl2019}
R.~C.~C. Leon, C.~H. Yang, J.~C.~C. Hwang, J.~Camirand Lemyre, T.~Tanttu,
  W.~Huang, K.~W. Chan, K.~Y. Tan, F.~E. Hudson, K.~M. Itoh, A.~Morello,
  A.~Laucht, M.~{Pioro-Ladriere}, A.~Saraiva, and A.~S. Dzurak.
\newblock Coherent spin control of s-, p-, d- and f-electrons in a silicon
  quantum dot.
\newblock \emph{arXiv:1902.01550 [cond-mat]}, February 2019.
\newblock URL \url{http://arxiv.org/abs/1902.01550}.

\bibitem[Itoh and Watanabe(2014)]{itohIsotopeEngineeringSilicon2014}
Kohei~M. Itoh and Hideyuki Watanabe.
\newblock Isotope engineering of silicon and diamond for quantum computing and
  sensing applications.
\newblock \emph{MRS Communications}, 4\penalty0 (4):\penalty0 143--157,
  December 2014.
\newblock \doi{10.1557/mrc.2014.32}.

\bibitem[Veldhorst et~al.(2015)Veldhorst, Yang, Hwang, Huang, Dehollain,
  Muhonen, Simmons, Laucht, Hudson, Itoh, Morello, and
  Dzurak]{veldhorstTwoqubitLogicGate2015}
M.~Veldhorst, C.~H. Yang, J.~C.~C. Hwang, W.~Huang, J.~P. Dehollain, J.~T.
  Muhonen, S.~Simmons, A.~Laucht, F.~E. Hudson, K.~M. Itoh, A.~Morello, and
  A.~S. Dzurak.
\newblock A two-qubit logic gate in silicon.
\newblock \emph{Nature}, 526\penalty0 (7573):\penalty0 410--414, October 2015.
\newblock \doi{10.1038/nature15263}.

\bibitem[Witzel et~al.(2015)Witzel, Monta{\~n}o, Muller, and
  Carroll]{witzelMultiqubitGatesProtected2015}
Wayne~M. Witzel, In{\`e}s Monta{\~n}o, Richard~P. Muller, and Malcolm~S.
  Carroll.
\newblock Multiqubit gates protected by adiabaticity and dynamical decoupling
  applicable to donor qubits in silicon.
\newblock \emph{Physical Review B}, 92\penalty0 (8):\penalty0 081407, August
  2015.
\newblock \doi{10.1103/PhysRevB.92.081407}.

\bibitem[Khaneja et~al.(2005)Khaneja, Reiss, Kehlet, {Schulte-Herbr{\"u}ggen},
  and Glaser]{khanejaOptimalControlCoupled2005}
Navin Khaneja, Timo Reiss, Cindie Kehlet, Thomas {Schulte-Herbr{\"u}ggen}, and
  Steffen~J. Glaser.
\newblock Optimal control of coupled spin dynamics: Design of {{NMR}} pulse
  sequences by gradient ascent algorithms.
\newblock \emph{Journal of Magnetic Resonance}, 172\penalty0 (2):\penalty0
  296--305, February 2005.
\newblock \doi{10.1016/j.jmr.2004.11.004}.

\bibitem[Yang et~al.(2012)Yang, Lim, Lai, Rossi, Morello, and
  Dzurak]{yangOrbitalValleyState2012}
C.~H. Yang, W.~H. Lim, N.~S. Lai, A.~Rossi, A.~Morello, and A.~S. Dzurak.
\newblock Orbital and valley state spectra of a few-electron silicon quantum
  dot.
\newblock \emph{Physical Review B}, 86\penalty0 (11):\penalty0 115319,
  September 2012.
\newblock \doi{10.1103/PhysRevB.86.115319}.

\bibitem[Ono et~al.(2002)Ono, Austing, Tokura, and
  Tarucha]{onoCurrentRectificationPauli2002}
K.~Ono, D.~G. Austing, Y.~Tokura, and S.~Tarucha.
\newblock Current {{Rectification}} by {{Pauli Exclusion}} in a {{Weakly
  Coupled Double Quantum Dot System}}.
\newblock \emph{Science}, 297\penalty0 (5585):\penalty0 1313--1317, August
  2002.
\newblock \doi{10.1126/science.1070958}.

\bibitem[Petta et~al.(2005)Petta, Johnson, Taylor, Laird, Yacoby, Lukin,
  Marcus, Hanson, and Gossard]{pettaCoherentManipulationCoupled2005}
J.~R. Petta, A.~C. Johnson, J.~M. Taylor, E.~A. Laird, A.~Yacoby, M.~D. Lukin,
  C.~M. Marcus, M.~P. Hanson, and A.~C. Gossard.
\newblock Coherent {{Manipulation}} of {{Coupled Electron Spins}} in
  {{Semiconductor Quantum Dots}}.
\newblock \emph{Science}, 309\penalty0 (5744):\penalty0 2180--2184, September
  2005.
\newblock \doi{10.1126/science.1116955}.

\bibitem[Johnson et~al.(2005{\natexlab{a}})Johnson, Petta, Marcus, Hanson, and
  Gossard]{johnsonSinglettripletSpinBlockade2005}
A.~C. Johnson, J.~R. Petta, C.~M. Marcus, M.~P. Hanson, and A.~C. Gossard.
\newblock Singlet-triplet spin blockade and charge sensing in a few-electron
  double quantum dot.
\newblock \emph{Physical Review B}, 72\penalty0 (16):\penalty0 165308, October
  2005{\natexlab{a}}.
\newblock \doi{10.1103/PhysRevB.72.165308}.

\bibitem[Betz et~al.(2015)Betz, Wacquez, Vinet, Jehl, Saraiva, Sanquer,
  Ferguson, and {Gonzalez-Zalba}]{betzDispersivelyDetectedPauli2015}
A.~C. Betz, R.~Wacquez, M.~Vinet, X.~Jehl, A.~L. Saraiva, M.~Sanquer, A.~J.
  Ferguson, and M.~F. {Gonzalez-Zalba}.
\newblock Dispersively {{Detected Pauli Spin}}-{{Blockade}} in a {{Silicon
  Nanowire Field}}-{{Effect Transistor}}.
\newblock \emph{Nano Letters}, 15\penalty0 (7):\penalty0 4622--4627, July 2015.
\newblock \doi{10.1021/acs.nanolett.5b01306}.

\bibitem[West et~al.(2019)West, Hensen, Jouan, Tanttu, Yang, Rossi,
  {Gonzalez-Zalba}, Hudson, Morello, Reilly, and
  Dzurak]{westGatebasedSingleshotReadout2019}
Anderson West, Bas Hensen, Alexis Jouan, Tuomo Tanttu, Chih-Hwan Yang,
  Alessandro Rossi, M.~Fernando {Gonzalez-Zalba}, Fay Hudson, Andrea Morello,
  David~J. Reilly, and Andrew~S. Dzurak.
\newblock Gate-based single-shot readout of spins in silicon.
\newblock \emph{Nature Nanotechnology}, page~1, March 2019.
\newblock \doi{10.1038/s41565-019-0400-7}.

\bibitem[Pakkiam et~al.(2018)Pakkiam, Timofeev, House, Hogg, Kobayashi, Koch,
  Rogge, and Simmons]{pakkiamSingleShotSingleGateRf2018}
P.~Pakkiam, A.~V. Timofeev, M.~G. House, M.~R. Hogg, T.~Kobayashi, M.~Koch,
  S.~Rogge, and M.~Y. Simmons.
\newblock Single-{{Shot Single}}-{{Gate}} rf {{Spin Readout}} in {{Silicon}}.
\newblock \emph{Physical Review X}, 8\penalty0 (4):\penalty0 041032, November
  2018.
\newblock \doi{10.1103/PhysRevX.8.041032}.

\bibitem[Yang et~al.(2011)Yang, Lim, Zwanenburg, and
  Dzurak]{yangDynamicallyControlledCharge2011}
C.~H. Yang, W.~H. Lim, F.~A. Zwanenburg, and A.~S. Dzurak.
\newblock Dynamically controlled charge sensing of a few-electron silicon
  quantum dot.
\newblock \emph{AIP Advances}, 1\penalty0 (4):\penalty0 042111, October 2011.
\newblock \doi{10.1063/1.3654496}.

\bibitem[Johnson et~al.(2005{\natexlab{b}})Johnson, Petta, Taylor, Yacoby,
  Lukin, Marcus, Hanson, and Gossard]{johnsonTripletSingletSpin2005}
A.~C. Johnson, J.~R. Petta, J.~M. Taylor, A.~Yacoby, M.~D. Lukin, C.~M. Marcus,
  M.~P. Hanson, and A.~C. Gossard.
\newblock Triplet\textendash{}singlet spin relaxation via nuclei in a double
  quantum dot.
\newblock \emph{Nature}, 435\penalty0 (7044):\penalty0 925--928, June
  2005{\natexlab{b}}.
\newblock \doi{10.1038/nature03815}.

\bibitem[Srinivasa et~al.(2015)Srinivasa, Xu, and
  Taylor]{srinivasaTunableSpinQubitCoupling2015}
V.~Srinivasa, H.~Xu, and J.~M. Taylor.
\newblock Tunable {{Spin}}-{{Qubit Coupling Mediated}} by a {{Multielectron
  Quantum Dot}}.
\newblock \emph{Physical Review Letters}, 114\penalty0 (22), June 2015.
\newblock \doi{10.1103/PhysRevLett.114.226803}.

\bibitem[Mehl et~al.(2014)Mehl, Bluhm, and
  DiVincenzo]{mehlTwoqubitCouplingsSinglettriplet2014}
Sebastian Mehl, Hendrik Bluhm, and David~P. DiVincenzo.
\newblock Two-qubit couplings of singlet-triplet qubits mediated by one quantum
  state.
\newblock \emph{Physical Review B}, 90\penalty0 (4), July 2014.
\newblock \doi{10.1103/PhysRevB.90.045404}.

\bibitem[{Harvey-Collard} et~al.(2017){Harvey-Collard}, Jacobson, Rudolph,
  Dominguez, Eyck, Wendt, Pluym, Gamble, Lilly, {Pioro-Ladri{\`e}re}, and
  Carroll]{harvey-collardCoherentCouplingQuantum2017}
Patrick {Harvey-Collard}, N.~Tobias Jacobson, Martin Rudolph, Jason Dominguez,
  Gregory A.~Ten Eyck, Joel~R. Wendt, Tammy Pluym, John~King Gamble, Michael~P.
  Lilly, Michel {Pioro-Ladri{\`e}re}, and Malcolm~S. Carroll.
\newblock Coherent coupling between a quantum dot and a donor in silicon.
\newblock \emph{Nature Communications}, 8\penalty0 (1):\penalty0 1029, October
  2017.
\newblock \doi{10.1038/s41467-017-01113-2}.

\bibitem[Reed et~al.(2016)Reed, Maune, Andrews, Borselli, Eng, Jura, Kiselev,
  Ladd, Merkel, Milosavljevic, Pritchett, Rakher, Ross, Schmitz, Smith, Wright,
  Gyure, and Hunter]{reedReducedSensitivityCharge2016}
M.~D. Reed, B.~M. Maune, R.~W. Andrews, M.~G. Borselli, K.~Eng, M.~P. Jura,
  A.~A. Kiselev, T.~D. Ladd, S.~T. Merkel, I.~Milosavljevic, E.~J. Pritchett,
  M.~T. Rakher, R.~S. Ross, A.~E. Schmitz, A.~Smith, J.~A. Wright, M.~F. Gyure,
  and A.~T. Hunter.
\newblock Reduced {{Sensitivity}} to {{Charge Noise}} in {{Semiconductor Spin
  Qubits}} via {{Symmetric Operation}}.
\newblock \emph{Physical Review Letters}, 116\penalty0 (11):\penalty0 110402,
  March 2016.
\newblock \doi{10.1103/PhysRevLett.116.110402}.

\bibitem[Loss and DiVincenzo(1998)]{lossQuantumComputationQuantum1998}
Daniel Loss and David~P. DiVincenzo.
\newblock Quantum computation with quantum dots.
\newblock \emph{Physical Review A}, 57\penalty0 (1):\penalty0 120, 1998.
\newblock \doi{10.1103/PhysRevA.57.120}.

\bibitem[Meunier et~al.(2011)Meunier, Calado, and
  Vandersypen]{meunierEfficientControlledphaseGate2011}
T.~Meunier, V.~E. Calado, and L.~M.~K. Vandersypen.
\newblock Efficient controlled-phase gate for single-spin qubits in quantum
  dots.
\newblock \emph{Physical Review B}, 83\penalty0 (12), March 2011.
\newblock \doi{10.1103/PhysRevB.83.121403}.

\bibitem[Watson et~al.(2018)Watson, Philips, Kawakami, Ward, Scarlino,
  Veldhorst, Savage, Lagally, Friesen, Coppersmith, Eriksson, and
  Vandersypen]{watsonProgrammableTwoqubitQuantum2018}
T.~F. Watson, S.~G.~J. Philips, E.~Kawakami, D.~R. Ward, P.~Scarlino,
  M.~Veldhorst, D.~E. Savage, M.~G. Lagally, Mark Friesen, S.~N. Coppersmith,
  M.~A. Eriksson, and L.~M.~K. Vandersypen.
\newblock A programmable two-qubit quantum processor in silicon.
\newblock \emph{Nature}, 555\penalty0 (7698):\penalty0 633--637, February 2018.
\newblock \doi{10.1038/nature25766}.

\bibitem[Huang et~al.(2019)Huang, Yang, Chan, Tanttu, Hensen, Leon, Fogarty,
  Hwang, Hudson, Itoh, Morello, Laucht, and
  Dzurak]{huangFidelityBenchmarksTwoqubit2019}
W.~Huang, C.~H. Yang, K.~W. Chan, T.~Tanttu, B.~Hensen, R.~C.~C. Leon, M.~A.
  Fogarty, J.~C.~C. Hwang, F.~E. Hudson, K.~M. Itoh, A.~Morello, A.~Laucht, and
  A.~S. Dzurak.
\newblock Fidelity benchmarks for two-qubit gates in silicon.
\newblock \emph{Nature}, 569\penalty0 (7757):\penalty0 532, May 2019.
\newblock \doi{10.1038/s41586-019-1197-0}.

\bibitem[Baart et~al.(2017)Baart, Fujita, Reichl, Wegscheider, and
  Vandersypen]{baartCoherentSpinexchangeQuantum2017}
Timothy~Alexander Baart, Takafumi Fujita, Christian Reichl, Werner Wegscheider,
  and Lieven Mark~Koenraad Vandersypen.
\newblock Coherent spin-exchange via a quantum mediator.
\newblock \emph{Nature Nanotechnology}, 12\penalty0 (1):\penalty0 26--30,
  January 2017.
\newblock \doi{10.1038/nnano.2016.188}.

\bibitem[Ferdous et~al.(2018)Ferdous, Chan, Veldhorst, Hwang, Yang,
  Sahasrabudhe, Klimeck, Morello, Dzurak, and
  Rahman]{ferdousInterfaceinducedSpinorbitInteraction2018}
Rifat Ferdous, Kok~W. Chan, Menno Veldhorst, J.~C.~C. Hwang, C.~H. Yang,
  Harshad Sahasrabudhe, Gerhard Klimeck, Andrea Morello, Andrew~S. Dzurak, and
  Rajib Rahman.
\newblock Interface-induced spin-orbit interaction in silicon quantum dots and
  prospects for scalability.
\newblock \emph{Physical Review B}, 97\penalty0 (24):\penalty0 241401, June
  2018.
\newblock \doi{10.1103/PhysRevB.97.241401}.

\bibitem[Tokura et~al.(2006)Tokura, {van der Wiel}, Obata, and
  Tarucha]{tokuraCoherentSingleElectron2006}
Yasuhiro Tokura, Wilfred~G. {van der Wiel}, Toshiaki Obata, and Seigo Tarucha.
\newblock Coherent {{Single Electron Spin Control}} in a {{Slanting Zeeman
  Field}}.
\newblock \emph{Physical Review Letters}, 96\penalty0 (4), January 2006.
\newblock \doi{10.1103/PhysRevLett.96.047202}.

\bibitem[Corna et~al.(2018)Corna, Bourdet, Maurand, Crippa, {Kotekar-Patil},
  Bohuslavskyi, Lavi{\'e}ville, Hutin, Barraud, Jehl, Vinet, Franceschi,
  Niquet, and Sanquer]{cornaElectricallyDrivenElectron2018}
Andrea Corna, L{\'e}o Bourdet, Romain Maurand, Alessandro Crippa, Dharmraj
  {Kotekar-Patil}, Heorhii Bohuslavskyi, Romain Lavi{\'e}ville, Louis Hutin,
  Sylvain Barraud, Xavier Jehl, Maud Vinet, Silvano~De Franceschi, Yann-Michel
  Niquet, and Marc Sanquer.
\newblock Electrically driven electron spin resonance mediated by
  spin\textendash{}valley\textendash{}orbit coupling in a silicon quantum dot.
\newblock \emph{npj Quantum Information}, 4\penalty0 (1):\penalty0 6, February
  2018.
\newblock \doi{10.1038/s41534-018-0059-1}.

\bibitem[Jock et~al.(2018)Jock, Jacobson, {Harvey-Collard}, Mounce, Srinivasa,
  Ward, Anderson, Manginell, Wendt, Rudolph, Pluym, Gamble, Baczewski, Witzel,
  and Carroll]{jockSiliconMetaloxidesemiconductorElectron2018}
Ryan~M. Jock, N.~Tobias Jacobson, Patrick {Harvey-Collard}, Andrew~M. Mounce,
  Vanita Srinivasa, Dan~R. Ward, John Anderson, Ron Manginell, Joel~R. Wendt,
  Martin Rudolph, Tammy Pluym, John~King Gamble, Andrew~D. Baczewski, Wayne~M.
  Witzel, and Malcolm~S. Carroll.
\newblock A silicon metal-oxide-semiconductor electron spin-orbit qubit.
\newblock \emph{Nature Communications}, 9\penalty0 (1):\penalty0 1768, May
  2018.
\newblock \doi{10.1038/s41467-018-04200-0}.

\bibitem[Tanttu et~al.(2019)Tanttu, Hensen, Chan, Yang, Huang, Fogarty, Hudson,
  Itoh, Culcer, Laucht, Morello, and
  Dzurak]{tanttuControllingSpinorbitInteractions2019}
Tuomo Tanttu, Bas Hensen, Kok~Wai Chan, Chih~Hwan Yang, Wister~Wei Huang,
  Michael Fogarty, Fay Hudson, Kohei Itoh, Dimitrie Culcer, Arne Laucht, Andrea
  Morello, and Andrew Dzurak.
\newblock Controlling {{Spin}}-{{Orbit Interactions}} in {{Silicon Quantum Dots
  Using Magnetic Field Direction}}.
\newblock \emph{Physical Review X}, 9\penalty0 (2):\penalty0 021028, May 2019.
\newblock \doi{10.1103/PhysRevX.9.021028}.

\bibitem[Elzerman et~al.(2004)Elzerman, Hanson, {Willems van Beveren}, Witkamp,
  Vandersypen, and Kouwenhoven]{elzermanSingleshotReadoutIndividual2004}
J.~M. Elzerman, R.~Hanson, L.~H. {Willems van Beveren}, B.~Witkamp, L.~M.~K.
  Vandersypen, and L.~P. Kouwenhoven.
\newblock Single-shot read-out of an individual electron spin in a quantum dot.
\newblock \emph{Nature}, 430\penalty0 (6998):\penalty0 431--435, July 2004.
\newblock \doi{10.1038/nature02693}.

\bibitem[{Harvey-Collard} et~al.(2018){Harvey-Collard}, D'Anjou, Rudolph,
  Jacobson, Dominguez, Ten~Eyck, Wendt, Pluym, Lilly, Coish,
  {Pioro-Ladri{\`e}re}, and
  Carroll]{harvey-collardHighFidelitySingleShotReadout2018}
Patrick {Harvey-Collard}, Benjamin D'Anjou, Martin Rudolph, N.~Tobias Jacobson,
  Jason Dominguez, Gregory~A. Ten~Eyck, Joel~R. Wendt, Tammy Pluym, Michael~P.
  Lilly, William~A. Coish, Michel {Pioro-Ladri{\`e}re}, and Malcolm~S. Carroll.
\newblock High-{{Fidelity Single}}-{{Shot Readout}} for a {{Spin Qubit}} via an
  {{Enhanced Latching Mechanism}}.
\newblock \emph{Physical Review X}, 8\penalty0 (2):\penalty0 021046, May 2018.
\newblock \doi{10.1103/PhysRevX.8.021046}.

\bibitem[Fogarty et~al.(2018)Fogarty, Chan, Hensen, Huang, Tanttu, Yang,
  Laucht, Veldhorst, Hudson, Itoh, Culcer, Ladd, Morello, and
  Dzurak]{fogartyIntegratedSiliconQubit2018}
M.~A. Fogarty, K.~W. Chan, B.~Hensen, W.~Huang, T.~Tanttu, C.~H. Yang,
  A.~Laucht, M.~Veldhorst, F.~E. Hudson, K.~M. Itoh, D.~Culcer, T.~D. Ladd,
  A.~Morello, and A.~S. Dzurak.
\newblock Integrated silicon qubit platform with single-spin addressability,
  exchange control and single-shot singlet-triplet readout.
\newblock \emph{Nature Communications}, 9\penalty0 (1):\penalty0 4370, October
  2018.
\newblock \doi{10.1038/s41467-018-06039-x}.

\bibitem[Wood and
  Gambetta(2018)]{woodQuantificationCharacterizationLeakage2018}
Christopher~J. Wood and Jay~M. Gambetta.
\newblock Quantification and characterization of leakage errors.
\newblock \emph{Physical Review A}, 97\penalty0 (3), March 2018.
\newblock \doi{10.1103/PhysRevA.97.032306}.

\bibitem[Preskill(1998)]{preskillFaulttolerantQuantumComputation1998}
John Preskill.
\newblock Fault-tolerant quantum computation.
\newblock In \emph{Introduction to {{Quantum Computation}} and
  {{Information}}}, pages 213--269. {WORLD SCIENTIFIC}, October 1998.
\newblock \doi{10.1142/9789812385253_0008}.

\bibitem[Gottesman()]{gottesmanStabilizerCodesQuantum}
Daniel Gottesman.
\newblock \emph{Stabilizer {{Codes}} and {{Quantum Error Correction}}}.
\newblock PhD thesis.

\bibitem[Aliferis and
  Terhal(2007)]{aliferisFaulttolerantQuantumComputation2007}
Panos Aliferis and Barbara~M. Terhal.
\newblock Fault-tolerant {{Quantum Computation}} for {{Local Leakage Faults}}.
\newblock \emph{Quantum Info. Comput.}, 7\penalty0 (1):\penalty0 139--156,
  January 2007.
\newblock URL \url{http://dl.acm.org/citation.cfm?id=2011706.2011715}.

\bibitem[Fowler(2013)]{fowlerCopingQubitLeakage2013}
Austin~G. Fowler.
\newblock Coping with qubit leakage in topological codes.
\newblock \emph{Physical Review A}, 88\penalty0 (4), October 2013.
\newblock \doi{10.1103/PhysRevA.88.042308}.

\bibitem[Suchara et~al.(2015)Suchara, Cross, and
  Gambetta]{sucharaLeakageSuppressionToric2015}
M.~Suchara, A.~W. Cross, and J.~M. Gambetta.
\newblock Leakage suppression in the toric code.
\newblock \emph{2015 IEEE International Symposium on Information Theory
  (ISIT)}, pages 1119--1123, June 2015.
\newblock \doi{10.1109/ISIT.2015.7282629}.

\bibitem[Barrett and Barnes(2002)]{barrettDoubleoccupationErrorsInduced2002}
S.~D. Barrett and C.~H.~W. Barnes.
\newblock Double-occupation errors induced by orbital dephasing in
  exchange-interaction quantum gates.
\newblock \emph{Physical Review B}, 66\penalty0 (12):\penalty0 125318,
  September 2002.
\newblock \doi{10.1103/PhysRevB.66.125318}.

\bibitem[Wang et~al.(2013)Wang, Payette, Dovzhenko, Deelman, and
  Petta]{wangChargeRelaxationSingleElectron2013}
K.~Wang, C.~Payette, Y.~Dovzhenko, P.~W. Deelman, and J.~R. Petta.
\newblock Charge {{Relaxation}} in a {{Single}}-{{Electron Si}} / {{SiGe Double
  Quantum Dot}}.
\newblock \emph{Physical Review Letters}, 111\penalty0 (4), July 2013.
\newblock \doi{10.1103/PhysRevLett.111.046801}.

\bibitem[Cai and Benjamin(2019)]{caiConstructingSmallerPauli2019}
Zhenyu Cai and Simon~C. Benjamin.
\newblock Constructing {{Smaller Pauli Twirling Sets}} for {{Arbitrary Error
  Channels}}.
\newblock \emph{Scientific Reports}, 9\penalty0 (1):\penalty0 1--11, August
  2019.
\newblock \doi{10.1038/s41598-019-46722-7}.

\bibitem[Gottesman(1998)]{gottesmanHeisenbergRepresentationQuantum1998}
Daniel Gottesman.
\newblock The {{Heisenberg Representation}} of {{Quantum Computers}}.
\newblock \emph{arXiv:quant-ph/9807006}, July 1998.
\newblock URL \url{http://arxiv.org/abs/quant-ph/9807006}.

\bibitem[Aaronson and
  Gottesman(2004)]{aaronsonImprovedSimulationStabilizer2004}
Scott Aaronson and Daniel Gottesman.
\newblock Improved simulation of stabilizer circuits.
\newblock \emph{Physical Review A}, 70\penalty0 (5):\penalty0 052328, November
  2004.
\newblock \doi{10.1103/PhysRevA.70.052328}.

\bibitem[Geller and Zhou(2013)]{gellerEfficientErrorModels2013}
Michael~R. Geller and Zhongyuan Zhou.
\newblock Efficient error models for fault-tolerant architectures and the
  {{Pauli}} twirling approximation.
\newblock \emph{Physical Review A}, 88\penalty0 (1), July 2013.
\newblock \doi{10.1103/PhysRevA.88.012314}.

\bibitem[Guti{\'e}rrez et~al.(2013)Guti{\'e}rrez, Svec, Vargo, and
  Brown]{gutierrezApproximationRealisticErrors2013}
Mauricio Guti{\'e}rrez, Lukas Svec, Alexander Vargo, and Kenneth~R. Brown.
\newblock Approximation of realistic errors by {{Clifford}} channels and
  {{Pauli}} measurements.
\newblock \emph{Physical Review A}, 87\penalty0 (3), March 2013.
\newblock \doi{10.1103/PhysRevA.87.030302}.

\bibitem[Guti{\'e}rrez and
  Brown(2015)]{gutierrezComparisonQuantumErrorcorrection2015}
Mauricio Guti{\'e}rrez and Kenneth~R. Brown.
\newblock Comparison of a quantum error-correction threshold for exact and
  approximate errors.
\newblock \emph{Physical Review A}, 91\penalty0 (2):\penalty0 022335, February
  2015.
\newblock \doi{10.1103/PhysRevA.91.022335}.

\bibitem[Stephens(2014)]{stephensFaulttolerantThresholdsQuantum2014}
Ashley~M. Stephens.
\newblock Fault-tolerant thresholds for quantum error correction with the
  surface code.
\newblock \emph{Physical Review A}, 89\penalty0 (2), February 2014.
\newblock \doi{10.1103/PhysRevA.89.022321}.

\bibitem[McKay et~al.(2017)McKay, Wood, Sheldon, Chow, and
  Gambetta]{mckayEfficientZGatesQuantum2017}
David~C. McKay, Christopher~J. Wood, Sarah Sheldon, Jerry~M. Chow, and Jay~M.
  Gambetta.
\newblock Efficient {{Z}}-{{Gates}} for {{Quantum Computing}}.
\newblock \emph{Physical Review A}, 96\penalty0 (2), August 2017.
\newblock \doi{10.1103/PhysRevA.96.022330}.

\bibitem[Zheng et~al.(2019)Zheng, Samkharadze, Noordam, Kalhor, Brousse,
  Sammak, Scappucci, and Vandersypen]{zhengRapidGatebasedSpin2019}
Guoji Zheng, Nodar Samkharadze, Marc~L. Noordam, Nima Kalhor, Delphine Brousse,
  Amir Sammak, Giordano Scappucci, and Lieven M.~K. Vandersypen.
\newblock Rapid gate-based spin read-out in silicon using an on-chip resonator.
\newblock \emph{Nature Nanotechnology}, 14\penalty0 (8):\penalty0 742--746,
  August 2019.
\newblock \doi{10.1038/s41565-019-0488-9}.

\bibitem[Raussendorf et~al.(2007)Raussendorf, Harrington, and
  Goyal]{raussendorfTopologicalFaulttoleranceCluster2007}
R.~Raussendorf, J.~Harrington, and K.~Goyal.
\newblock Topological fault-tolerance in cluster state quantum computation.
\newblock \emph{New Journal of Physics}, 9\penalty0 (6):\penalty0 199, 2007.
\newblock \doi{10.1088/1367-2630/9/6/199}.

\bibitem[Hwang et~al.(2017)Hwang, Yang, Veldhorst, Hendrickx, Fogarty, Huang,
  Hudson, Morello, and Dzurak]{hwangImpactFactorsValleys2017}
J.~C.~C. Hwang, C.~H. Yang, M.~Veldhorst, N.~Hendrickx, M.~A. Fogarty,
  W.~Huang, F.~E. Hudson, A.~Morello, and A.~S. Dzurak.
\newblock Impact of \$g\$-factors and valleys on spin qubits in a silicon
  double quantum dot.
\newblock \emph{Physical Review B}, 96\penalty0 (4):\penalty0 045302, July
  2017.
\newblock \doi{10.1103/PhysRevB.96.045302}.

\bibitem[Malinowski et~al.(2018)Malinowski, Martins, Smith, Bartlett, Doherty,
  Nissen, Fallahi, Gardner, Manfra, Marcus, and
  Kuemmeth]{malinowskiSpinMultielectronQuantum2018}
Filip~K. Malinowski, Frederico Martins, Thomas~B. Smith, Stephen~D. Bartlett,
  Andrew~C. Doherty, Peter~D. Nissen, Saeed Fallahi, Geoffrey~C. Gardner,
  Michael~J. Manfra, Charles~M. Marcus, and Ferdinand Kuemmeth.
\newblock Spin of a {{Multielectron Quantum Dot}} and {{Its Interaction}} with
  a {{Neighboring Electron}}.
\newblock \emph{Physical Review X}, 8\penalty0 (1):\penalty0 011045, March
  2018.
\newblock \doi{10.1103/PhysRevX.8.011045}.

\bibitem[Kolmogorov(2009)]{kolmogorovBlossomNewImplementation2009}
Vladimir Kolmogorov.
\newblock Blossom {{V}}: A new implementation of a minimum cost perfect
  matching algorithm.
\newblock \emph{Mathematical Programming Computation}, 1\penalty0 (1):\penalty0
  43--67, July 2009.
\newblock \doi{10.1007/s12532-009-0002-8}.

\bibitem[Htt()]{HttpsGithubCom}
{{https://github.com/czydbb/SurfaceCodeModule}}.
\newblock URL \url{https://github.com/czydbb/SurfaceCodeModule}.

\bibitem[Bravyi et~al.(2014)Bravyi, Suchara, and
  Vargo]{bravyiEfficientAlgorithmsMaximum2014}
Sergey Bravyi, Martin Suchara, and Alexander Vargo.
\newblock Efficient algorithms for maximum likelihood decoding in the surface
  code.
\newblock \emph{Physical Review A}, 90\penalty0 (3), September 2014.
\newblock \doi{10.1103/PhysRevA.90.032326}.

\end{thebibliography}

\end{document}